\documentclass[10pt,journal]{IEEEtran}
\usepackage{cite}
\usepackage{amsmath,amssymb,amsfonts}
\usepackage{algorithmic}
\usepackage[ruled,longend,linesnumbered]{algorithm2e}
\usepackage{graphicx}
\usepackage{textcomp}
\usepackage{xcolor}
\def\BibTeX{{\rm B\kern-.05em{\sc i\kern-.025em b}\kern-.08em
    T\kern-.1667em\lower.7ex\hbox{E}\kern-.125emX}}

\usepackage{indentfirst} 
\usepackage{comment}   
\usepackage{multirow}  
\usepackage{booktabs}  
\usepackage{multirow}  
\usepackage{array}     
\usepackage{enumitem}

\usepackage{subcaption} 
\usepackage{tabularx}
\usepackage{multirow}
\usepackage{makecell}

\usepackage{url}

\begin{document}

\title{Revealing the Potential and Challenges of LLMs for Wireless Network Fault Diagnosis: A Multi-Modal Wi-Fi Benchmark}

\author{
	Junjian Zhang,
	Haobo Deng,
	Xinxin Li,
	Ming Zhao,~\IEEEmembership{Member,~IEEE},
	Fengxiao Tang,~\IEEEmembership{Senior Member,~IEEE},
	Nei Kato,~\IEEEmembership{Fellow,~IEEE}%
	
	\thanks{This work was supported in part by the Hunan Provincial Natural Science Foundation under Grant 2025JJ90177, and in part by the Jiangxi Provincial Natural Science Foundation under Grant 20253BAC280098.}%
	\thanks{Junjian Zhang, Haobo Deng, Xinxin Li, and Ming Zhao are with Central South University, Changsha, China.}
	\thanks{Fengxiao Tang is with Central South University, Changsha, China, and also with Tohoku University, Sendai 980-8576, Japan (corresponding author; e-mail: tangfengxiao@csu.edu.cn).}
	\thanks{Nei Kato is with Tohoku University, Sendai 980-8576, Japan.}
}

\maketitle

\begin{abstract}

Intelligent network operation and maintenance systems in modern networks continuously generate large volumes of multi-modal operational data.
However, the correspondence between wireless network fault behaviors and multi-modal operational
data remains insufficiently studied, as isolated and fragmented evaluations make it difficult to systematically compare.
We establish a unified benchmark for multi-modal Wi-Fi fault diagnosis to systematically compare diagnosis performance across tasks, modalities, and representative paradigms.
We construct WITNESS, a real-world multi-modal Wi-Fi fault dataset collected from a wireless testbed with automated fault injection, comprising over 10,000 samples.
To the best of our knowledge, this is among the first publicly available datasets jointly capturing heterogeneous cross-layer operational observations for Wi-Fi fault diagnosis.
Benchmark analysis reveals task-dependent modality effects and the limited ability of conventional methods to exploit multi-modal operational data.
Beyond prediction-oriented evaluation, we further examine emerging LLM-based approaches, and develop a reasoning-oriented evaluation framework to assess whether generated diagnostic analyses remain consistent with actual network conditions.
Our analysis shows that LLM-based approaches exhibit potential in generating diagnostic reasoning, but still face challenges in capturing fine-grained numerical patterns in operational data.
Overall, these findings suggest several important considerations for future multi-modal wireless network fault diagnosis.

\end{abstract}


\section{Introduction}

Reliable Wi-Fi fault diagnosis is becoming a fundamental capability for intelligent network operation and maintenance in modern wireless systems, where learning-based techniques have been widely applied to address growing network scale, device heterogeneity, and operational complexity~\cite{FL_wifi, Intelligent_network_1, Intelligent_network_2}.
At the same time, modern networks continuously generate multi-modal operational data~\cite{Intelligent_network_data_1, Intelligent_network_data_2}, providing complementary evidence for diagnosing complex failures, as increasingly explored in AIOps and microservice diagnosis~\cite{medicine2024, zhang2025microservice, aiops_reliability_llm}.
In Wi-Fi fault diagnosis, however, studies still largely focus on individual operational modalities, leaving the role of cross-layer multi-modal observations insufficiently understood~\cite{dasilva2022packetloss, aiops2023}.

The first challenge lies in obtaining real-world multi-modal Wi-Fi fault observations, which are difficult to collect without stable and reproducible Wi-Fi fault injection mechanisms.
Public wireless datasets for Wi-Fi sensing~\cite{bravenec2023uji} and traffic analysis~\cite{mawi_archive} generally lack network anomaly behaviors.
Reproducible anomaly datasets such as AWID3~\cite{chatzoglou2021awid3}, ToN-IoT~\cite{moustafa2020toniotwindows}, and IoT-23~\cite{garcia2020iot23} mainly focus on intrusion or security anomalies rather than operational network faults.
Meanwhile, real-world Wi-Fi operational datasets collected from enterprise networks~\cite{witals,agv_wifi_dmesg} often contain sparse fault observations and are rarely publicly available.
This makes it difficult to characterize Wi-Fi fault behaviors and fairly compare diagnosis methods across modalities and tasks.

Second, Wi-Fi fault diagnosis lacks systematic analysis of multi-modal operational data.
Although additional modalities often improve diagnosis accuracy, aggregate performance is insufficient to characterize modality-specific contributions across fault types and diagnosis tasks.
In AIOps and microservice diagnosis, traces, logs, metrics, and KPIs have been jointly used to analyze failures and examine modality contributions~\cite{eadro,diagfusion2023,medicine2024}.
Such analyses are usually task-specific, leaving cross-fault modality-level analysis insufficiently explored.
In Wi-Fi fault diagnosis, systematic modality-level analysis remains particularly limited.
Existing studies often focusing on specific observational views, such as AP-side metrics~\cite{witals}, packet/radio traces~\cite{jigsaw}, packet-loss and ACK-pattern behaviors~\cite{wislow}, or predefined wireless pathology symptoms~\cite{wifi_pathology_llm}.
This limits the ability of diagnosis algorithms to fully exploit multi-modal advantages.

LLMs provide a new opportunity for multi-modal Wi-Fi fault diagnosis by leveraging prior knowledge, rather than relying solely on data-driven feature learning.
A third challenge therefore lies in evaluating LLM-based Wi-Fi fault diagnosis.
LLM-based diagnosis often feeds outputs into downstream models~\cite{netllm,network_semantic_llm,zhang2025aiops,cloud_rca_llm}, accurate final predictions do not necessarily indicate accurate LLM-generated analyses.
Meanwhile, existing LLM-output evaluation methods are not well suited to Wi-Fi fault diagnosis.
Human-annotated reference-based metrics~\cite{zhang2019bertscore} are subjective and may overlook fine-grained operational differences across fault instances.
Reference-based evaluations using rule-derived~\cite{llm_rule} are not directly applicable to network fault diagnosis.
This calls for an evaluation framework beyond prediction-oriented metrics to assess the intrinsic diagnostic capability of LLMs for network diagnosis.

To provide a reproducible basis for studying multi-modal Wi-Fi fault diagnosis, we construct a real-world Wi-Fi fault diagnosis testbed with controlled fault injection in campus environments.
We organize the selected faults according to the Wi-Fi message delivery process, covering representative operational layers that span software execution, traffic congestion, association management, MAC-layer transmission, and hardware conditions.
Using this testbed, we collect a multi-modal dataset\cite{WITNESS} containing over 10,000 samples across diverse wireless scenarios, jointly capturing multi-modal operational data under representative Wi-Fi fault conditions.
Considering the modality coverage of both public network datasets~\cite{CICIoT2023, garcia2020iot23,moustafa2020toniotwindows} and private operational datasets~\cite{witals,agv_wifi_dmesg}, we select four modalities, including metrics, traces, events, and logs.
We establish a unified benchmark.\footnote{\url{https://github.com/CSUFUNLAB/WITNESS}} for multi-modal Wi-Fi fault diagnosis across representative methods.
The benchmark supports analysis beyond overall diagnosis accuracy by examining how effectively different methods leverage multiple operational modalities, revealing fault-specific differences that are obscured by aggregate metrics.

Beyond prediction-oriented evaluation, we further examine LLM-generated diagnostic analyses, where models provide open-ended fault interpretations in addition to final predictions.
To make such analyses quantitatively assessable, we derive reference operational features from underlying observations using network-fault-specific rules, avoiding the need for human-written diagnostic references.
We measure their consistency and evaluate whether LLM explanations are grounded in actual network conditions rather than merely plausible text.
Through this evaluation, we investigate the relationship among LLM prior knowledge, reasoning consistency, and final diagnosis performance in multi-modal Wi-Fi fault diagnosis.
The results reveal both the potential and challenges of LLMs for multi-modal Wi-Fi fault diagnosis.

Our main contributions are summarized as follows:

\begin{itemize}
	
\item We provide, to the best of our knowledge, the first publicly available multi-modal Wi-Fi fault diagnosis dataset, offering a reproducible basis for studying Wi-Fi fault behaviors under unified evaluation settings.

\item We conduct an empirical analysis of multi-modal Wi-Fi fault diagnosis, examining how different operational modalities contribute to diagnosis under different Wi-Fi fault conditions.

\item We develop a structured reasoning evaluation methodology for Wi-Fi fault diagnosis, enabling quantitative assessment of whether LLM-generated diagnostic analyses reflect underlying operational conditions.

\end{itemize}

The rest of this paper is organized as follows. Section II reviews related work. Section III presents our multi-modal Wi-Fi fault diagnosis benchmark. Section IV introduces the reasoning-oriented evaluation framework for LLM-based fault analysis. Section V presents the experimental results and empirical analysis. Finally, Section VI concludes this paper.

\section{Related Work}

\subsection{Real-world Data for Wi-Fi Fault Diagnosis}

Recent Wi-Fi studies increasingly rely on real-world measurements rather than purely simulation-based analysis to capture practical wireless behaviors~\cite{WFI_CHANNEL, NB_IOT, Semi_sim}.
Public wireless datasets have provided valuable resources for studying Wi-Fi sensing~\cite{WiFi_sensing1, WiFi_sensing2, bravenec2023uji,hanna2022wisig}, signal-based applications\cite{WiFi_Signals}, traffic analysis~\cite{mawi_archive}, intrusion detection~\cite{chatzoglou2021awid3,sharafaldin2018toward}, and IoT anomaly analysis~\cite{garcia2020iot23,moustafa2020toniotwindows,moustafa2020toniotlinux}.
CIC-series datasets further provide realistic testbeds and benchmark resources for intrusion detection and security analysis in IoT, IoV, and IoMT environments~\cite{CICIoV2024,CICIoMT2024,CICIoT2022}.
However, these datasets are usually designed for sensing, security, or traffic characterization tasks, rather than operational Wi-Fi fault diagnosis.
They therefore provide limited support for analyzing how specific Wi-Fi faults are reflected across synchronized cross-layer observations.

Existing Wi-Fi diagnosis and troubleshooting studies further provide important insights into practical wireless problems.
WiSlow~\cite{wislow} targets end-user Wi-Fi performance troubleshooting, Witals~\cite{witals} performs AP-centric health diagnosis of Wi-Fi networks, and packet-level or cross-layer studies analyze packet loss and anomaly behaviors in Wi-Fi or wireless community networks~\cite{dasilva2022packetloss,wcn_fault_detection}.
Recent work also investigates diagnosis in specific industrial scenarios, such as AGV Wi-Fi disconnection analysis using dmesg logs~\cite{agv_wifi_dmesg}.
These datasets and studies provide useful resources for understanding wireless traffic, sensing behaviors, security threats, or practical Wi-Fi performance problems.
However, most of them are not designed as Wi-Fi fault diagnosis datasets, Wi-Fi fault diagnosis still lacks a common benchmark basis for studying.

\subsection{Multi-Modal Failure Diagnosis}

Multi-modal failure diagnosis has been increasingly studied in AIOps, microservice systems, and distributed software environments.
These systems commonly rely on heterogeneous observability data, including logs, metrics, traces, and KPIs, to diagnose failures and identify root causes.
Eadro~\cite{eadro} studies end-to-end troubleshooting for microservices using multi-source data, DiagFusion~\cite{diagfusion2023} performs robust failure diagnosis through multi-modal data, and MEDICINE~\cite{medicine2024} investigates modality contribution through multi-modal adaptive optimization.
Related surveys and benchmark datasets further show the importance of multi-source observability for failure diagnosis and root cause analysis~\cite{zhang2025microservice,multisource_aiops_data,zheng2024lemmarca,pham2024rcaeval}.

Although these studies demonstrate the value of multi-modal diagnosis, their system models and fault mechanisms differ substantially from Wi-Fi networks.
Microservice and AIOps systems mainly involve software-level dependencies, service calls, logs, metrics, and traces, whereas Wi-Fi faults are shaped by wireless channel conditions, MAC-layer behaviors, device heterogeneity, traffic dynamics, and host-side system states.
Therefore, conclusions drawn from AIOps or microservice benchmarks cannot directly characterize the role of multi-modal operational data in Wi-Fi fault diagnosis.
For Wi-Fi networks, how different modalities contribute to fault detection, classification, and localization remains insufficiently understood.

\subsection{LLM-Based Network Diagnosis and Evaluation}

Large language models have recently been adapted to networking, system operation, and fault diagnosis tasks\cite{LLM_Management}.
NetLLM~\cite{netllm} explores adapting large language models for networking, while recent surveys summarize the potential of LLMs for communication, network, and service management~\cite{llm_network_management_survey,zhang2025aiops}.
LLM-based methods have also been studied for cloud incident root cause analysis~\cite{cloud_rca_llm}, reliability evaluation under chaos engineering settings~\cite{aiops_reliability_llm}, and log-based anomaly detection with LLM-enhanced representations~\cite{llmelog_feature}.
Compared with conventional classifiers, these approaches can generate natural-language diagnostic analyses in addition to final fault predictions.

This capability also introduces new evaluation challenges.
Prediction-oriented metrics can evaluate whether the final fault label is correct, but they cannot determine whether the generated diagnostic analysis reflects the underlying operational conditions.
Reference-based text similarity metrics, such as BERTScore~\cite{zhang2019bertscore}, can compare generated text with reference descriptions, but they are less suitable for verifying sample-specific consistency between diagnostic interpretations and network observations.
This limitation is particularly important for multi-modal Wi-Fi fault diagnosis, where similar fault descriptions may correspond to different quantitative behaviors across performance metrics, packet traces, warning events, and system logs.
Thus, evaluating LLM-based Wi-Fi diagnosis requires reasoning-level metrics that are grounded in operational network states rather than only final predictions or textual similarity.

\section{Multi-modal Wi-Fi Fault Diagnosis Benchmark}
\label{sec:benchmark}

Our objective is to establish a unified benchmark for Wi-Fi fault diagnosis.
In practical network operations, fault data often rely on manual inspection or post-hoc annotation to identify fault events and construct diagnosis labels, making them difficult to collect and reproduce at scale.
The key challenge in building such a benchmark is therefore to construct fault scenarios that are both realistic and controllable.
We develop an automated fault injection and data collection system deployed across a central control platform and node-side agents, enabling repeatable fault injection, recovery control, traffic coordination, and synchronized multi-modal data collection.
Table~\ref{benchmark_Scale} summarizes the scale statistics of our benchmark.

\begin{table}[!t]
	\centering
	\caption{Scale Statistics of Wi-Fi Fault Diagnosis Benchmark}
	\label{benchmark_Scale}
	\renewcommand{\arraystretch}{1.15}
	\begin{tabular}{ll}
		\toprule
		\textbf{Aspect} & \textbf{Scale} \\
		\midrule
		Device nodes & 21 \\
		Wi-Fi standards & 3 \\
		Network scenarios & 3 \\
		Operational modalities & 4 \\
		Fault types & 11 \\
		Baseline methods & 13 \\
		LLM families & 4 \\
		\bottomrule
	\end{tabular}
\end{table}

\begin{figure}[t]
	\centering
	\includegraphics[width=\linewidth]{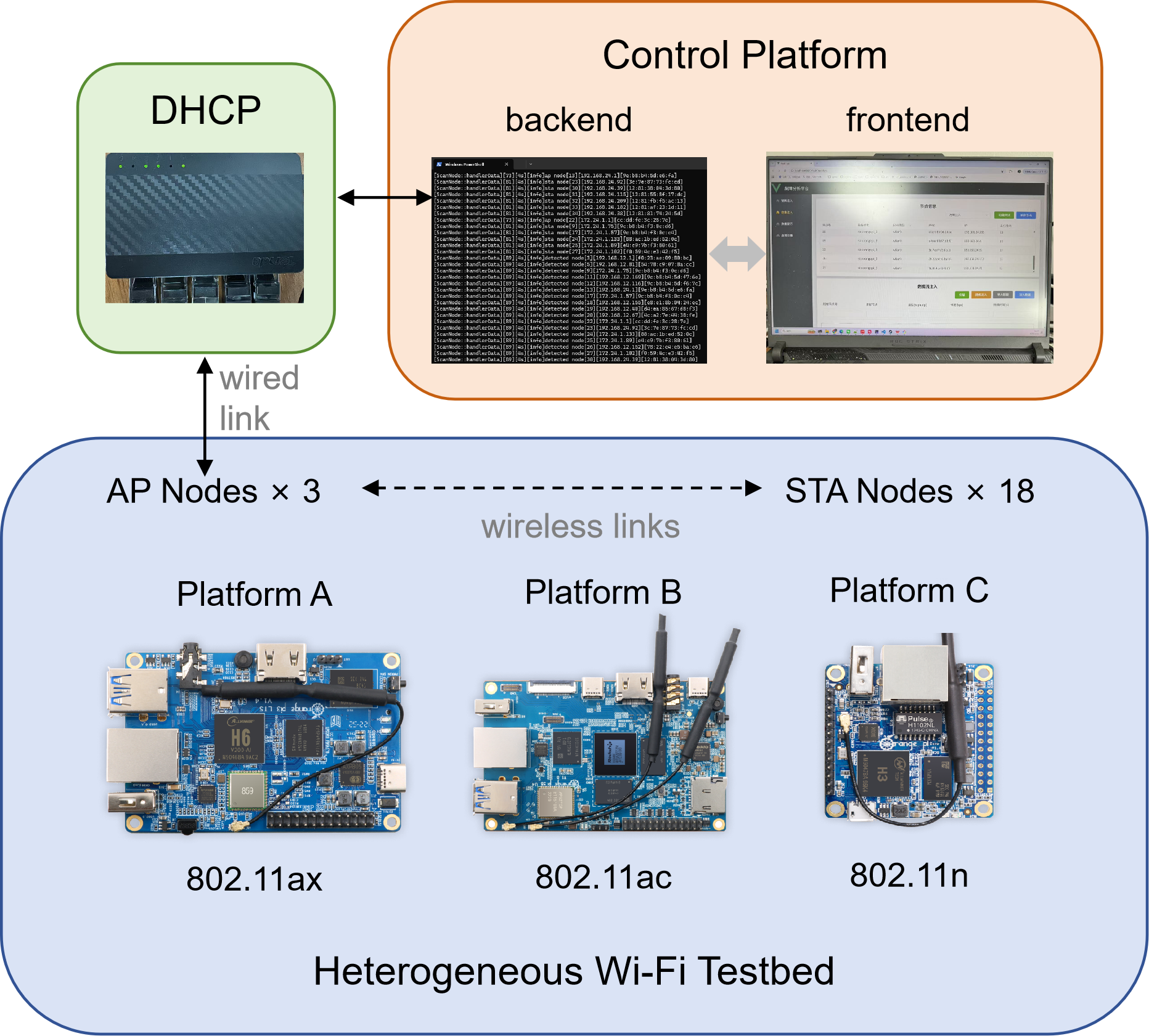}
	\caption{Overview of the Wi-Fi Fault Inject Testbed.}
	\label{fig:testbed}
\end{figure}

\subsection{Wi-Fi Fault Injection Testbed}

The testbed consists of two main components: a centralized control platform and node-side agents.
The centralized control platform configures and coordinates individual wireless nodes through network commands.
Under fault conditions, packet loss and delay may make centralized control commands unreliable.
Node-side agents are deployed to ensure that each node executes scheduled operations locally, including traffic generation, fault injection, and automatic recovery.

Fig.~\ref{fig:testbed} illustrates the overview of our fault inject Wi-Fi testbed.
To cover representative Wi-Fi operating conditions, the testbed incorporates diversity in hardware platforms, network deployments, and traffic patterns.
At the hardware level, it consists of 21 Wi-Fi nodes with three heterogeneous hardware and software configurations spanning IEEE 802.11n/ac/ax wireless capabilities.
Table~\ref{tab:testbed_config} summarizes the hardware and software configurations used in the testbed.
At the deployment level, it covers both infrastructure-based and ad hoc\cite{adhoc} wireless networks with H2H and IoT traffic patterns.
To better approximate realistic network conditions, we use public network datasets to construct traffic matrices.
Specifically, we construct the following three workload scenarios, excluding H2H ad hoc scenarios because they are uncommon in practical Wi-Fi deployments:

\begin{itemize}
	\item H2H AP--STA scenario: MAWI traffic traces from the WIDE Project in Japan~\cite{H2h_Pub}.
	
	\item IoT AP--STA scenario: IoT-23 traffic traces from CTU University in the Czech Republic~\cite{garcia2020iot23}.
	
	\item IoT ad hoc scenario: CRAWDAD umass/diesel traces from UMassDieselNet in the United States~\cite{adhoc_pub}.
	
\end{itemize}

Traffic matrices alone cannot fully characterize workload dynamics, and we model H2H traffic with heavy-tailed characteristics to capture bursty human-driven communication\cite{heavy_traffic}, while modeling IoT traffic with light-tailed characteristics to reflect relatively regular and lightweight device communication\cite{light_traffic}.

\begin{figure}[t]
	\centering
	\begin{subfigure}{0.48\linewidth}
		\centering
		\includegraphics[width=\linewidth]{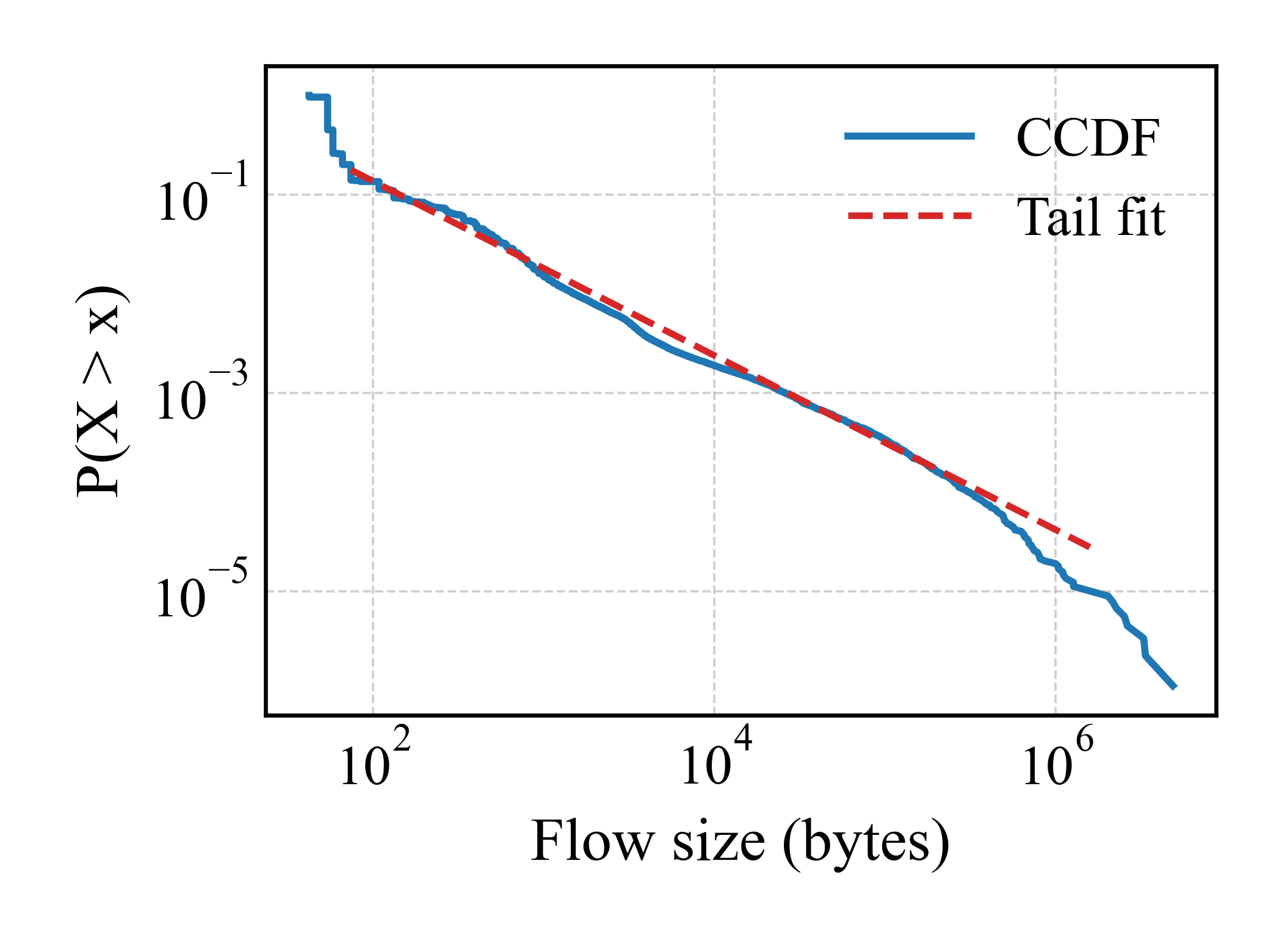}
		\caption{H2H, log-log scale.}
	\end{subfigure}
	\hfill
	\begin{subfigure}{0.48\linewidth}
		\centering
		\includegraphics[width=\linewidth]{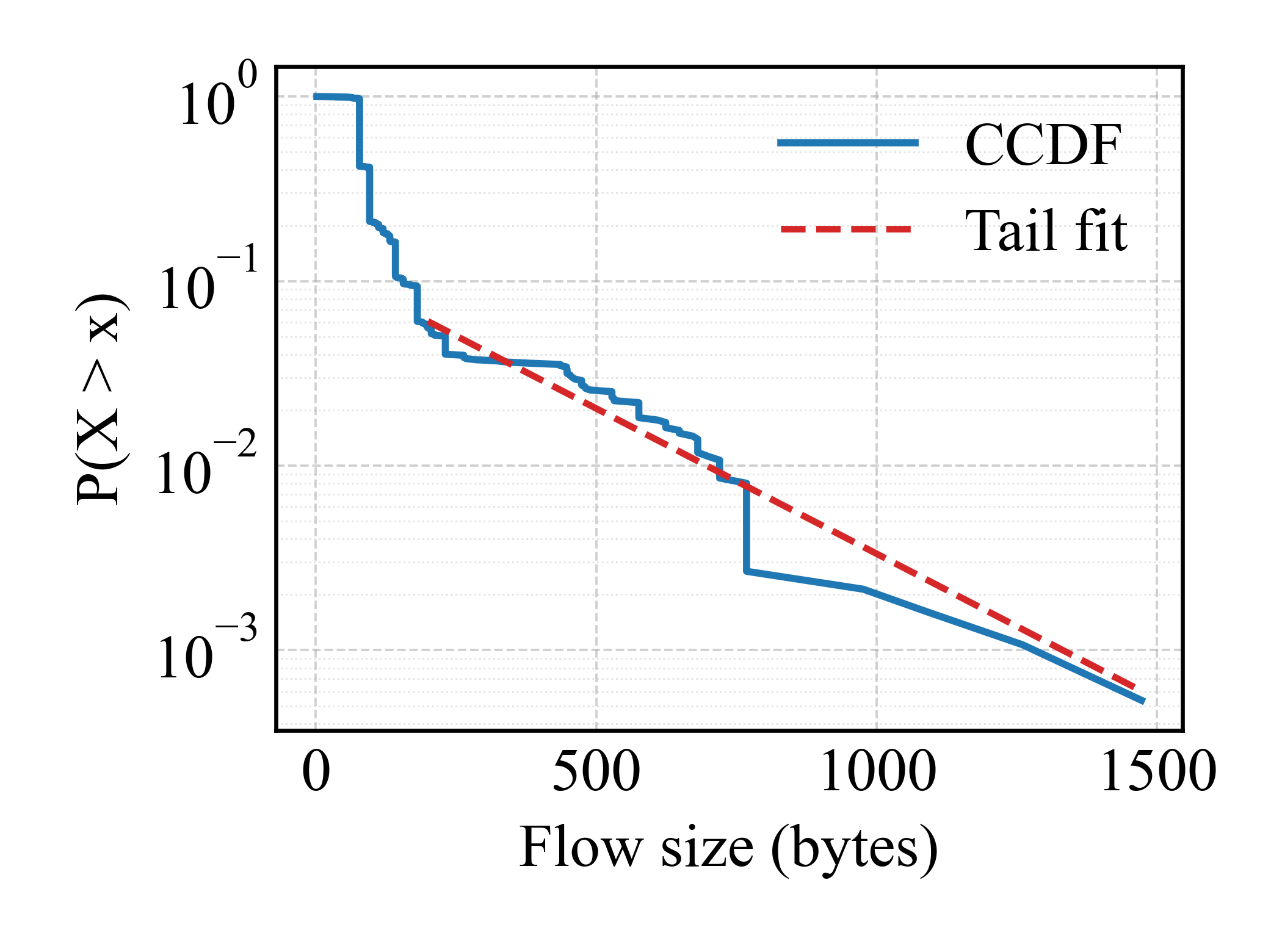}
		\caption{IoT semi-log scale.}
	\end{subfigure}
	\caption{Flow-size CCDF analysis for H2H and IoT traffic.}
	\label{fig:ccdf_h2h_iot}
\end{figure}

We use the empirical complementary cumulative distribution function (CCDF) of traffic sizes in the selected public traces to validate their tail characteristics as shown in Fig~\ref{fig:ccdf_h2h_iot}.
The H2H traces show an approximately linear tail in the log-log CCDF plot, and the IoT traces show an approximately linear decay in the semi-log CCDF plot. Based on these observations, we use Pareto distributions for H2H and Poisson distributions for IoT to reproduce the corresponding workload characteristics.

\subsection{Fault Scenario Construction}

Wi-Fi fault diagnosis covers a broad problem space, making it infeasible to reproduce all possible faults in a single benchmark.
Therefore, considering both representativeness and repeatability, we select 11 representative fault types across five categories.
These categories reflect key stages of Wi-Fi message delivery, ranging from software execution and traffic congestion to association management, MAC-layer transmission, and hardware conditions.
Table~\ref{fault_table} summarizes the selected fault types, their categories, sample distributions, and descriptions.
It is worth noting that the fault set differs slightly in ad hoc scenarios due to their decentralized communication structure.
Since there is no access point, Probe Failure and Beacon Loss are represented by Link Failure.
Rate Adaptation Failure is not included due to hardware limitations.
Hidden Node is also not included, because our ad hoc topology is configured to keep all nodes directly reachable, preventing hidden-terminal conditions.

\subsection{Multi-modal Data Collection}

Considering the modality coverage of existing network datasets~\cite{mawi_archive, garcia2020iot23,moustafa2020toniotwindows,moustafa2020toniotlinux, multisource_aiops_data} and the complementary information provided by different operational views, we collect four types of multi-modal data: performance metrics, traces, events, and logs.
These modalities are commonly generated in modern network operation and maintenance systems, and together capture Wi-Fi fault behaviors
Performance metrics, including throughput, latency, jitter, and packet loss, are collected using iperf for traffic generation.

Packet traces are collected on each node using tcpdump, which is a common practice in public network datasets.
Raw pcap files are usually large and contain substantial low-level details that are redundant for sample-level fault diagnosis.
We further process packet traces using CICFlowMeter\cite{CICFlowMeter} to extract flow-level statistical features.
Although this preprocessing simplifies raw packet traces, it also serves as a feature extraction step that improves information density.

Thirteen types of warning events are generated by an additional warning system.
In practice, such event data are highly customized, lack unified standards, and are closely tied to operator or vendor products.
Our lightweight warning system could monitors common network metrics and runtime states,  but the generated events do not directly correspond to specific faults.
This reflects a key characteristic of warning events: they are originally designed as error-code-like signals for specific abnormalities, but real network behaviors often go beyond their original design assumptions.

Six types of runtime monitoring logs are collected using built-in Linux utilities, instead of system logs, as system logs cannot reflect the injected Wi-Fi fault behaviors.
We select runtime indicators that are closely related to fault symptoms, such as CPU utilization, memory usage, process states, traffic volume statistics, and signal strength indicators.
We do not further transform these plain-text monitoring records into serialized feature sequences.
Instead, we preserve their textual form to allow LLM-based methods to directly exploit the original runtime descriptions.

All modalities are collected within the same observation window, ensuring that they correspond to the same fault instance.
Each sample is collected within a 3-minute observation window.
The first minute records normal network operation, after which the fault is injected.
For faults with configurable severity, the fault reaches its strongest state around the second minute.

\begin{table}[!t]
	\centering
	\caption{Testbed Configurations}
	\label{tab:testbed_config}
	\begin{tabular}{llll}
		\toprule
		\textbf{Wi-Fi Chipset} & \textbf{Ubuntu OS / Kernel} & \textbf{Wi-Fi Capability} \\
		\midrule
		AP6275P & Jammy / Linux 6.1 & IEEE 802.11ax \\
		AW859A & Focal / linux 5.10 & IEEE 802.11ac \\
		XR819 & Bionic / Linux 5.4 & IEEE 802.11n \\
		\bottomrule
	\end{tabular}
\end{table}

\begin{table*}[!t]
	\centering
	\caption{Summary of Fault Types and Sample Distribution}
	\label{fault_table}
	\renewcommand{\arraystretch}{1.25}
	\begin{tabularx}{\textwidth}{
			>{\centering\arraybackslash}m{2.5cm}
			>{\centering\arraybackslash}m{2.5cm}
			>{\centering\arraybackslash}m{2cm}
			>{\raggedright\arraybackslash}X
		}
		\toprule
		\textbf{Fault Type} & \makecell[c]{\textbf{Samples} \\ (H2H / IoT / ad hoc)} & \textbf{Fault Category} & \makecell[c]{\textbf{Description}} \\
		\midrule
		Normal
		& 1635 / 1724 / 1065 
		& -
		& Normal running \\
		Node Crash
		& 326 / 361 / 200 
		& Hardware  
		& A node becomes unresponsive due to system or hardware failure, leading to complete communication loss. \\
		Poor Link Quality
		& 334 / 421 / 200
		& Hardware
		& Degraded wireless link conditions (e.g., low RSSI or high interference) result in increased packet loss and unstable throughput. \\
		Application Crash
		& 424 / 357 / 200 
		& Software 
		& Application process terminates unexpectedly, reducing service availability and causing communication interruption.. \\
		Application Slowdown
		& 380 / 359 / 200 
		& Software 
		& Application level failures or performance degradation reduce service availability or significantly increase response latency. \\
		Rate Adaptation Failure
		& 226 / 205 / 0 
		& MAC-layer
		& Inefficient or incorrect rate selection causes suboptimal throughput and increased retransmissions under varying channel conditions. \\
		Hidden Node
		& 377 / 330 / 0
		& MAC-layer
		& Simultaneous transmissions from nodes outside each other's sensing range cause collisions at the receiver. \\
		Traffic Overload
		& 384 / 351 / 201 
		& Congestion
		& Excessive traffic demand exceeds network capacity, leading to congestion, delay, and packet loss. \\
		Buffer Bloat
		& 259 / 343 / 200 
		& Congestion
		& Excessive buffering increases queueing delay, leading to high latency without necessarily causing immediate packet drops. \\
		Queue Overflow
		& 256 / 340 / 203 
		& Congestion
		& Queue saturation causes packets to be dropped when incoming traffic exceeds the available buffer capacity. \\
		Probe Failure
		& 304 / 440 / 0
		& Association
		& Missing or failed probe responses prevent nodes from discovering or establishing connectivity with the target network. \\
		Beacon Loss
		& 226 / 194 / 0
		& Association
		& Missing beacon frames disrupt connectivity maintenance, causing unstable association or disconnection. \\
		Link Failure
		& 0 / 0 / 200
		& Association
		& The node cannot access the ad hoc network due to link failure, leading to communication loss. \\
		\bottomrule
	\end{tabularx}
\end{table*}

\subsection{Evaluation Setup}

Our baselines consider three typical fault diagnosis tasks, including fault detection, classification, and localization.
Fault detection is formulated as a binary classification task between normal and abnormal network conditions.
Fault classification is an 11-class classification task over fault types after excluding normal samples.
Fault localization is formulated as a node-level multi-class classification task for identifying the affected node.
Precision, Recall, and F1-score are used to evaluate the three tasks.
We evaluate representative methods from three diagnosis paradigms: traditional machine learning, time-series deep learning, and LLM-based diagnosis.
Traditional methods include DT, RF, SVM, KNN, and MLP, while time-series deep learning methods include LSTM, CNN, and GRU.
For LLM-based diagnosis, we adopt a feature extraction approach with representative LLM families, including GPT, Gemini, Qwen, and LLaMA, since direct end-to-end diagnosis with LLMs shows ineffective prediction performance.
In practice, fully LLM-based diagnosis often relies on a more complex diagnostic pipeline\cite{llm_pipeline} rather than a single direct LLM inference step.

Then, we preprocess the collected data for benchmark evaluation and provide both the raw data and the preprocessed files. The common preprocessing steps are summarized as follows:

\begin{itemize}
		
	\item Normalization:
	All numerical features are normalized using min-max normalization.
	
	\item Node Anonymization:
	Node indices are anonymized during preprocessing by randomly permuting node ordering within each sample to prevent diagnosis models from exploiting topology-specific behaviors associated with fixed node identities.

	\item Trace Feature filtering:
	CICFlowMeter produces a large number of statistical features from packet traces.
	To avoid giving packet traces an unfair advantage over other modalities, we retain only averaged statistical features for benchmark evaluation.
\end{itemize}

For traditional machine learning methods, we compute the mean value of each feature dimension for every node and aggregate the node-level features into fixed-size sample representations.
For time-series deep learning methods, variable-length time-series observations are truncated or padded to a predefined length determined by the average sequence length in the dataset.
For LLM-based methods, numerical observations are converted into z-score deviation levels relative to normal samples, mitigating the numerical insensitivity of LLMs.

\section{Reasoning-oriented Evaluation Framework}
\label{sec:Reasoning-oriented Evaluation}


\subsection{Overview}

We develop a reasoning-oriented evaluation framework to assess whether LLM-generated diagnostic analyses are grounded in actual operational conditions.
Unlike prediction-oriented evaluation, which only measures final fault labels, this framework evaluates the diagnostic evidence expressed in open-ended LLM outputs.

\begin{figure}[t]
	\centering
	\includegraphics[width=\linewidth]{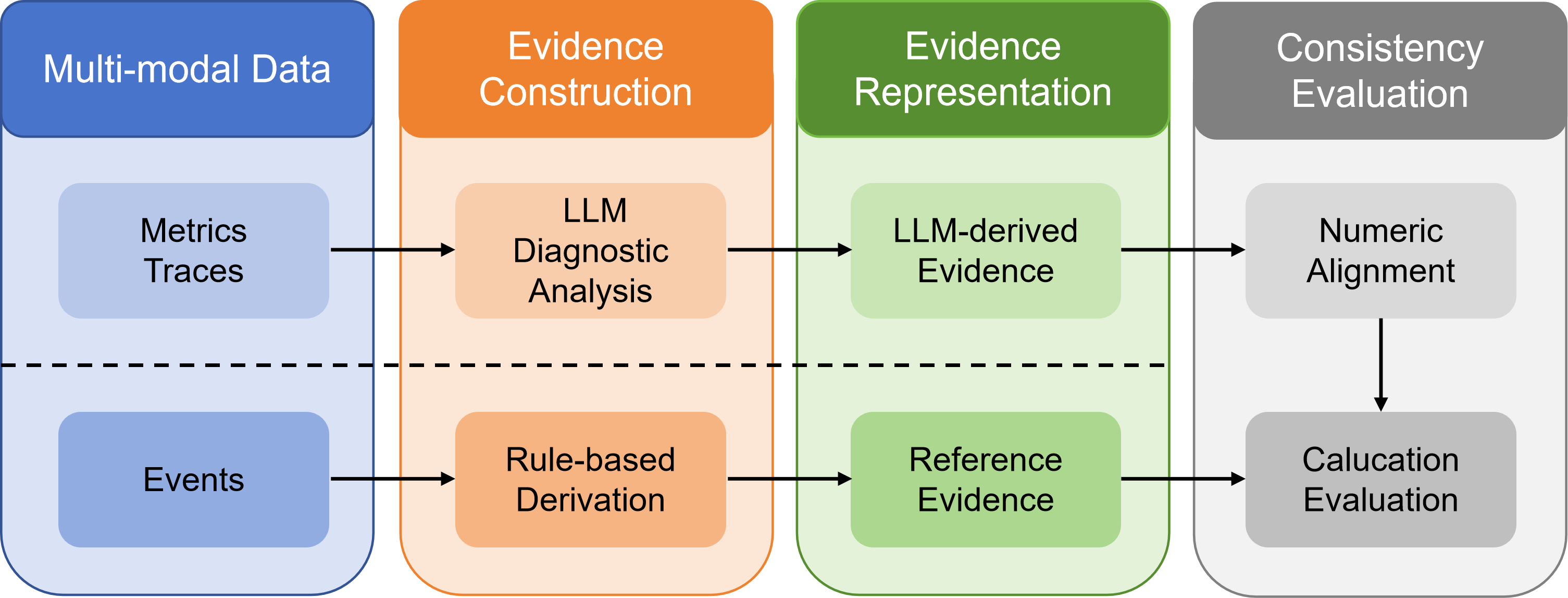}
	\caption{reasoning-oriented evaluation framework Overview.}
	\label{reasoning_framework}
\end{figure}

The overall workflow is illustrated in Fig.~\ref{reasoning_framework}.
Given a sample with multi-modal operational observations, we first prompt the LLM to generate a diagnostic analysis and extract a structured diagnostic evidence representation from it.
In parallel, we derive a reference evidence representation from the original data using rule-based mappings.
The consistency between the two evidence representations is then measured to quantify whether the LLM analysis reflects the actual network state.

Two design issues are critical for this evaluation.
First, the LLM analysis path and the reference derivation path should not rely on identical input modalities; otherwise, the evaluation may degenerate into testing whether the LLM reproduces deterministic rules.
Second, LLM-derived and data-derived evidence may follow different value distributions because they are produced through different construction processes, so an alignment step is required before consistency measurement.
The framework does not assume a fixed evidence form, and can be instantiated using feature-based evidence vectors, causal chains, or other structured representations.
In this study, we instantiate the evidence representation as network state features and use binarization for alignment.

\subsection{Network State Feature}

In this implementation, we instantiate the diagnostic evidence representation as network state features score.
These features can be obtained by modifying the prompt to require structured supporting evidence from LLM outputs.
The score-based design captures the varying severity of the same fault under different scenarios, rather than assigning identical evidence templates to all samples with the same fault label.
They are particularly suitable for feature-extraction-based LLM diagnosis pipelines, because the extracted state features can be directly used by downstream diagnosis models.

It is worth noting that warning events indicate specific runtime events detected by the monitoring system, such as a threshold violation or the occurrence of a particular signal.
In contrast, network state features provide score-based descriptions of overall network behaviors, such as system load, link quality, traffic congestion, or network performance degradation.
Therefore, warning events capture what explicitly happened at a certain moment, whereas network state features summarize how the network behaves over the observation window.

We select ten network state features that reflect fault-induced operational conditions, while keeping them sufficiently general to avoid one-to-one correspondence with specific fault types.
These features cover symptom-level conditions, network-side communication quality, application-side runtime status, and system-side operating states.
Each feature is represented by a normalized value ranging from 0 to 1, as summarized below.

\begin{itemize}
	\item Stutter symptom: unstable communication reflected by throughput fluctuation, increased delay, or jitter without complete interruption.
	
	\item Interruption symptom: communication interruption reflected by connectivity loss, zero throughput, or complete transmission failure.
	
	\item Delay severity: abnormal transmission delay reflected by increased RTT, latency, or delay-related warnings.
	
	\item Packet loss severity: packet delivery loss reflected by packet loss statistics or packet-loss-related warnings.
	
	\item Jitter severity: delay variation over time reflected by unstable packet latency.
	
	\item Poor signal quality: degraded wireless signal quality reflected by weak RSSI, low SNR, or signal-related warnings.
	
	\item Hardware issue: hardware-side abnormalities such as device crashes, overheating, memory/CPU failures, or hardware warnings.
	
	\item Network issue: network communication problems including delay, packet loss, jitter, throughput instability, or connection interruption.
	
	\item Application issue: application-level failures such as service crashes, abnormal responses, or application-related warnings.
	
	\item High system load: system load pressure reflected by abnormal CPU usage or load-related warnings.
\end{itemize}

\subsection{Reference State Feature Derivation}

Using rule-based derivation for reference state feature derivation avoids the need for additional manual annotation\cite{llmelog_feature}.
This enables automatic and scalable reference construction with consistent criteria across samples.
In general, deriving network states from raw operational data requires complex rule design.
Fortunately, our dataset includes warning events, which provide structured signals and simplify this process.
We directly map warning events and fault labels to network state features.

Let \(W\) denotes warning event observations, \(F\) denotes fault type annotations.
The mapping function \(f(\cdot)\) encodes predefined associations between warning conditions and fault type.
\(E_{\mathrm{data}}\) represents the resulting operational feature representation, which is defined as:

\begin{equation}
\begin{aligned}
E^{data} &= [e_1, e_2, \ldots, e_n] \\
e_i &= \min(1, \ e_i^{W} + e_i^{F}) \\
e_i^{W} &= \frac{\sum_{w_j \in W} f_i^{W}(w_j)}{f_{i,\max}^{W}}, \\
e_i^{F} &= f_i^{F}(F)
\end{aligned}
\end{equation}
The overall procedure first maps warning events and fault types to operational semantic features through predefined mapping tables \(f^{W}\) and \(f^{F}\).
The fault-derived scores $e_i^{F}$ are deterministic, where each fault type is directly mapped to a binary semantic feature vector.
The warning-derived scores $e_i^{W}$ are summed over all warning events and then normalized.
The final data-derived score $E^{data}$ is obtained by summing the warning-derived and fault-derived scores, capped at 1.
Thus, $E^{data}$ is a hybrid operational semantic representation that combines deterministic fault-derived semantics with normalized warning-derived evidence scores.
More importantly, it reflects the actual node-level operational conditions, rather than simply providing a label-based reference template.

LLM-derived feature scores can be obtained through prompting and reflect the model's interpretation of network conditions.
They are produced under a different scoring mechanism from $E^{data}$, and thus are not directly comparable to the data-derived evidence scores.
We apply binary mapping to convert them into stricter semantic activations, where each activation indicates whether the corresponding operational semantic feature is identified by the LLM-generated analysis.
Let $e^{LLM} \in [0,1]^d$ denote the operational feature vector generated by the LLM. To enable feature alignment, the LLM-generated feature vector is binarized using a threshold vector \(\boldsymbol{\tau} \in [0,1]^d\), where each dimension has its own threshold. The binary operational feature vector \(E_{\mathrm{llm}}\) is defined as:

\begin{equation}
	E^{LLM}_{i} =
	\begin{cases}
		1, & e_{i}^{LLM} \ge \tau_{i} \\
		0, & \text{otherwise}
	\end{cases}
\end{equation}

The reasoning consistency is evaluated using Weighted Explanation Precision (WEP), Weighted Explanation Recall (WER), and Weighted Explanation F1 (WEF1), defined as:

\begin{equation}
	\begin{aligned}
		\mathrm{WEP} &=
		\frac{
			\sum_{i=1}^{n} E_i^{\mathrm{LLM}} E_i^{\mathrm{data}}
		}{
			\sum_{i=1}^{n} E_i^{\mathrm{LLM}}
		}, \\ \\
		\mathrm{WER} &=
		\frac{
			\sum_{i=1}^{n} E_i^{\mathrm{LLM}} E_i^{\mathrm{data}}
		}{
			\sum_{i=1}^{n} E_i^{\mathrm{data}}
		}, \\  \\
		\mathrm{WEF1} &=
		\frac{2 \cdot \mathrm{WEP} \cdot \mathrm{WER}}
		{\mathrm{WEP} + \mathrm{WER}}.
	\end{aligned}
\end{equation}
The $E_{\mathrm{llm}}^{(i)}$ threshold $\tau^{(i)}$ is treated as a tunable parameter and is selected to maximize the WEF1 on the training data. Since different operational features may exhibit different output distributions under LLM-based analysis, each feature dimension is allowed to adopt an independent threshold for feature activation.

This yields an automated evidence-level evaluation that reduces ambiguity in free-form text comparison and quantifies the consistency between LLM-generated explanations and actual network states.

\section{Results and Multi-modal Analysis}

\subsection{Overall Performance}

We first present the overall experimental results in Table~\ref{all_fault_result}. 
The table reports the F1-scores of different diagnosis methods across three observation modalities: iperf metrics, tcpdump traces, and warning events. 
Each entry is shown as \((D/C/L)\), denoting fault detection, classification, and localization, respectively.
The results indicate that the dataset is neither too easy, where simple methods can solve all tasks, nor too noisy, where no method can extract useful fault-related information. 
Therefore, the benchmark provides a reasonable basis for further analyzing modality effectiveness, task difficulty, and model behavior.
%
%
We do not evaluate traditional methods on monitor logs, as this requires additional log serialization and feature engineering, while also introducing potential overlap with warning-event features.
Given their naturally textual form, monitor logs are instead used for LLM-based diagnostic analysis.

We provide a modality-level overview of the results, and further discuss modality effects in the subsequent task- and method-oriented analyses.
As shown in Table~\ref{all_fault_result}, the diagnostic F1-scores generally increase from iperf metrics to tcpdump traces and warning events.
Warning events naturally achieve the strongest performance, as they provide high-level abstractions of underlying network states.
Contrary to the intuitive expectation, iperf metrics, which more directly reflect network performance, achieve lower diagnostic performance than tcpdump traces.
From a human observer's perspective, increased delay and packet loss in iperf metrics provide direct evidence of abnormal network conditions.
CICFlowMeter-derived trace features do not directly report network performance metrics, but capture traffic-level behaviors related to performance degradation.
For example, delay- and jitter-related behaviors can be reflected by timing features such as IAT-related features, which capture variations in packet arrival intervals and flow completion time.
Loss-related degradation is not directly measured, but may be reflected by changes in packet and byte rates, bidirectional traffic statistics and Down/Up Ratio.
Moreover, TCP flag features can further expose abnormal connection behaviors.
Although the performance implications behind these statistics are not directly intuitive to human observers, machine learning models can exploit their latent patterns for fault diagnosis.
With additional flow-level statistics, these trace-based features can provide more discriminative information than iperf metrics.

At a high level, the observed performance trend aligns with the increasing information density of the three modalities.
However, the F1-score is only an aggregate metric and cannot fully characterize the detailed behavior of fault diagnosis. 
We therefore further examine how different modalities affect fault classification in the following analysis.

\begin{table}[!t]
	\centering
	\caption{Overall diagnosis performance across modalities and tasks.}
	\label{all_fault_result}
	\begin{tabular}{cccc}
		\toprule
		\textbf{Method} & \textbf{Metrics} (D/C/L) & \textbf{Traces} (D/C/L) & \textbf{Events} (D/C/L)\\
		\midrule
		LR  & 0.77 / 0.46 / 0.31  & 0.81 / 0.62 / 0.46  & 0.87 / 0.80 / 0.68 \\
		MLP  & 0.74 / 0.42 / 0.32  & 0.80 / 0.64 / 0.50  & 0.84 / 0.81 / 0.69 \\
		RF  & 0.68 / 0.43 / 0.33  & 0.82 / 0.61 / 0.46  & 0.74 / 0.68 / 0.61 \\
		SVM  & 0.68 / 0.41 / 0.30  & 0.81 / 0.62 / 0.45  & 0.83 / 0.77 / 0.62 \\
		KNN  & 0.73 / 0.34 / 0.29  & 0.80 / 0.51 / 0.37  & 0.84 / 0.74 / 0.58 \\
		\midrule		
		CNN  & 0.69 / 0.34 / 0.41  & 0.87 / 0.58 / 0.36  & 0.74 / 0.31 / 0.32 \\
		GRU  & 0.79 / 0.44 / 0.42  & 0.88 / 0.66 / 0.46  & 0.69 / 0.47 / 0.30 \\
		LSTM  & 0.78 / 0.43 / 0.39  & 0.86 / 0.60 / 0.39  & 0.75 / 0.46 / 0.33 \\
		\midrule
		gemini  & 0.76 / 0.44 / 0.31  & 0.80 / 0.43 / 0.31  & 0.87 / 0.70 / 0.66 \\
		gpt  & 0.78 / 0.42 / 0.30  & 0.79 / 0.38 / 0.27  & 0.83 / 0.64 / 0.57 \\
		llama  & 0.76 / 0.37 / 0.25  & 0.79 / 0.39 / 0.23  & 0.80 / 0.47 / 0.54 \\
		qwen  & 0.73 / 0.24 / 0.22  & 0.77 / 0.31 / 0.23  & 0.74 / 0.27 / 0.29 \\
		\bottomrule
	\end{tabular}
	\vspace{0.5em}
\end{table}

\subsection{Method-level Analysis}

\begin{figure}[t]
	\centering
	\includegraphics[width=\linewidth]{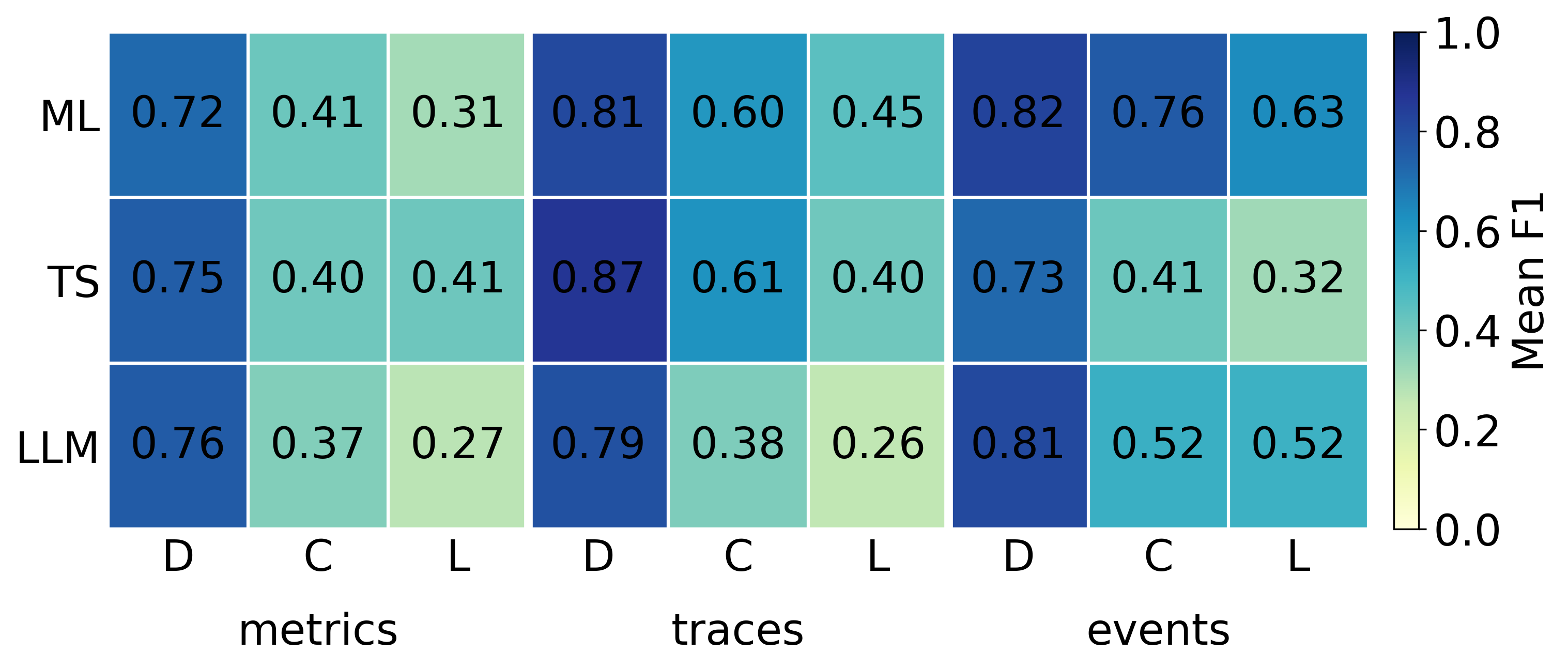}
	\caption{Mean F1 by Method Category, Modality, and Task.}
	\label{method_setting_heatmap}
\end{figure}

Since methods within the same category show broadly similar performance, except for a few outliers, we group the methods into traditional machine learning (ML), time-series models (TS), and LLM-based approaches (LLM). 
For ease of comparison, we report the average performance of methods within each category, as shown in heatmap Fig.~\ref{method_setting_heatmap}.
The ML category achieves the best performance across most settings, indicating that these methods remain strong baselines for this benchmark.
Although the original data are time-series in nature, the TS category does not achieve strong performance in our benchmark.
While TS models are designed to learn local sequential patterns, state transitions, and temporal dependencies from observations, these capabilities do not necessarily imply that they can uncover the fault semantics behind such temporal variations.
The limitation is reflected in the event modality: TS methods perform worse on Events than on Metrics and Traces, whereas ML and LLM-based methods show the opposite trend.
This is because the temporal evolution of warning events carries limited meaning in our setting; the occurrence of specific events is more important than their ordering.

For LLM-based methods, the situation is different.
The relatively weak performance is expected, since these models are evaluated without task-specific fine-tuning, their diagnosis capability largely depends on whether their pre-trained knowledge can correctly interpret our fault scenarios.
The results suggest that prompting alone is insufficient to enable LLMs to fully capture the complex semantics of our fault scenarios.
The text-analysis capability of LLMs is clearly reflected in the event modality, where they achieve performance close to traditional methods without additional task-specific configuration.
Meanwhile, smaller LLMs such as Llama and Qwen perform much worse than larger proprietary models such as Gemini and GPT
This observation is natural but also concerning. 
Larger proprietary models usually require additional paid access, introducing an unexpected cost compared with conventional diagnosis methods.

To further investigate the ability of LLMs to understand network states, we adopt an embedding-based evaluation approach. 
The results are shown in Table~\ref{LLM-embedding}.
The Llama2 embedding model is unofficial and yields relatively weak performance, whereas the Qwen embedding model brings a noticeable improvement.
Compared with direct prompting, embedding-based evaluation removes one layer of generation and format constraints.
The improvement suggests that, LLMs may encode more network-state semantics than what is expressed through the prompt-constrained feature outputs.
It also sacrifices the textual output capability of LLMs, namely their ability to provide natural-language diagnostic analyses of faults.

\begin{table}[!t]
	\centering
	\caption{Diagnosis performance of LLM embedding-based methods.}
	\label{LLM-embedding}
	\begin{tabular}{cccc}
		\toprule
		\textbf{Method} & \textbf{Metrics} (D/C/L) & \textbf{Traces} (D/C/L) & \textbf{Events} (D/C/L)\\
		\midrule
		llama2  & 0.81 / 0.19 / 0.16  & 0.81 / 0.32 / 0.13  & 0.81 / 0.24 / 0.22 \\
		qwen  & 0.81 / 0.37 / 0.16  & 0.82 / 0.45 / 0.14  & 0.79 / 0.53 / 0.37 \\
		\bottomrule
	\end{tabular}
	\vspace{0.5em}
\end{table}

Finally, we evaluate LLM-based fault diagnosis using monitor logs.
The results are even weaker than those obtained from other modalities.
A closer inspection of the LLM-generated analyses shows that the model can hardly distinguish normal logs from problematic logs in this setting.
One important reason is that monitor logs are still largely driven by numerical variables rather than purely semantic descriptions.
Although they are presented in textual form, many log messages follow fixed templates, where the informative parts are numerical values or parameter changes.
From an engineering perspective, encoding fixed log messages into event codes is often more direct and reliable.
For formatted string logs, the variable fields are also frequently numerical, which means that log analysis still suffers from the LLM's limited sensitivity to fine-grained quantitative variations.
This result further suggests that textual form alone does not guarantee that a modality is naturally suitable for LLM-based diagnosis.

The experimental results show that advanced methods do not naturally exhibit advantages on network operational data, as their capabilities do not automatically adapt to raw network observations.
This usually requires additional data processing steps.

\subsection{Task-level Analysis}

\begin{table}[!t]
	\centering
	\caption{End-to-end fault classification.}
	\label{End_classification}
	\begin{tabular}{cccc}
		\toprule
		\textbf{Method} & \textbf{Metrics} (P/R/F) & \textbf{Traces} (P/R/F) & \textbf{Events} (P/R/F)\\
		\midrule
		LR  & 0.52 / 0.40 / 0.43  & 0.62 / 0.49 / 0.53  & 0.80 / 0.67 / 0.70 \\
		MLP  & 0.51 / 0.37 / 0.41  & 0.67 / 0.45 / 0.52  & 0.84 / 0.61 / 0.67 \\
		RF  & 0.52 / 0.35 / 0.37  & 0.64 / 0.52 / 0.55  & 0.68 / 0.45 / 0.51 \\
		SVM  & 0.51 / 0.36 / 0.38  & 0.70 / 0.54 / 0.57  & 0.78 / 0.61 / 0.65 \\
        KNN  & 0.40 / 0.27 / 0.32  & 0.59 / 0.44 / 0.49  & 0.77 / 0.62 / 0.67 \\
		\bottomrule
	\end{tabular}
	\vspace{0.5em}
\end{table}

Across the three diagnostic tasks, performance generally decreases from fault detection to classification and further to localization, indicating an increasing level of diagnostic difficulty.
It should be noted that fault classification is evaluated conditionally on faulty samples.
Table~\ref{End_classification} reports the end-to-end classification performance using precision, recall, and F1-score, denoted as \((P/R/F)\). 
A prediction is considered correct only if the sample is first detected as faulty and its fault type is subsequently classified correctly.
Normal samples are excluded from this evaluation, as their large proportion may otherwise inflate the overall accuracy.
%
A noteworthy observation is that introducing the detection stage does not always reduce the final classification F1-score. 
This is because the detection stage changes the error distribution of fault classification. 
For fault types that originally suffer from many cross-fault false positives, the detection stage may filter out part of these misclassified samples as normal, thereby reducing false positives. 
Although this may also introduce false negatives by rejecting some faulty samples, the resulting precision-recall trade-off can still lead to a higher F1-score for certain fault types.
Overall, this composite task serves as the practical classification pipeline, while our baseline analysis focuses on the individual diagnostic tasks rather than the combined pipeline.

\begin{figure}[t]
	\centering
	\includegraphics[width=\linewidth]{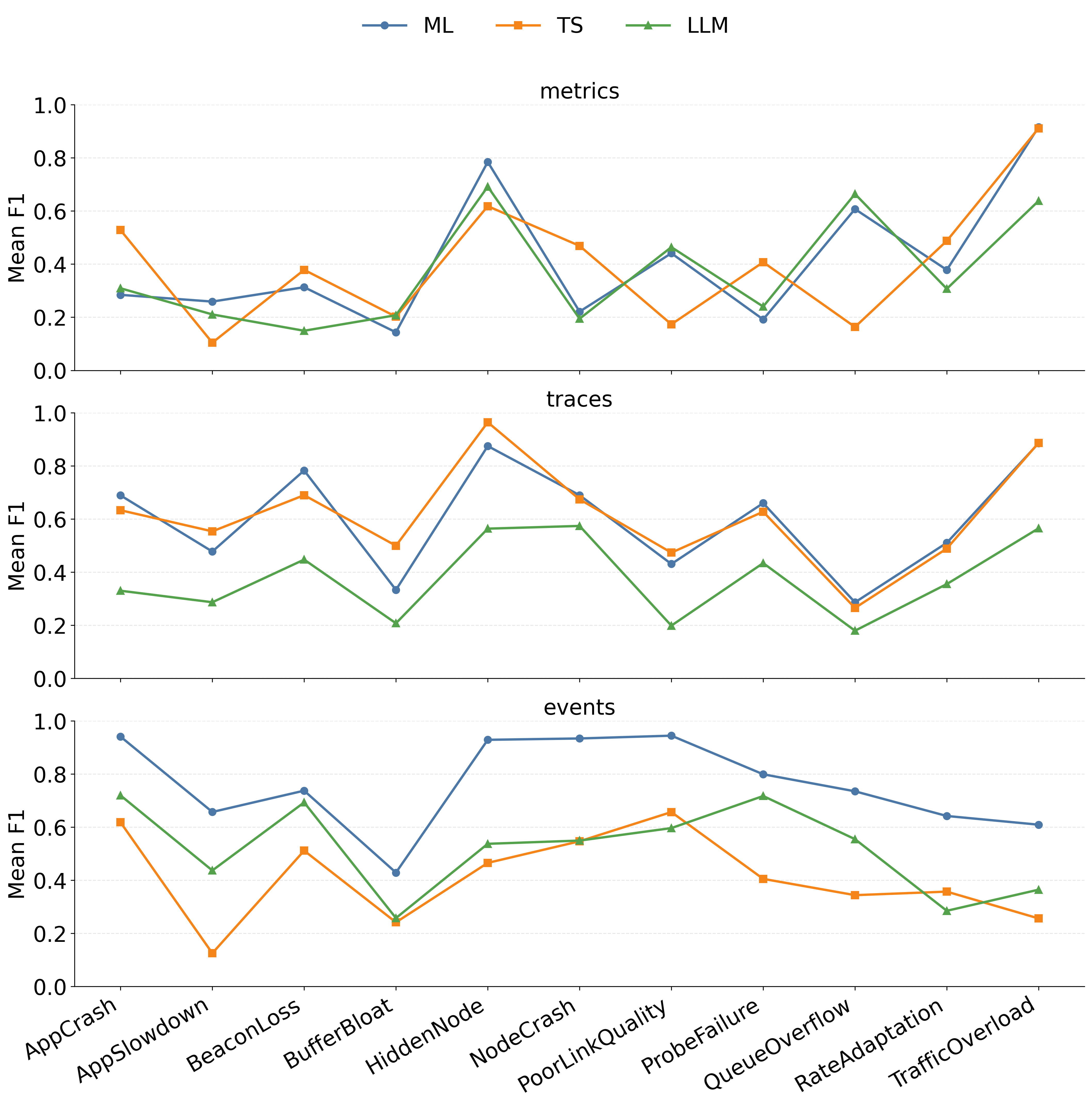}
	\caption{Per-Fault Classification F1 Across Method Categories and Modalities.}
	\label{method_category_classification_f1_lines}
\end{figure}

The previous Table~\ref{all_fault_result} reports only the aggregate F1-score for fault classification, which cannot characterize the modality-specific effects on individual fault types.
We decompose the fault classification task and analyze the results at the fault type level. 
For each modality, we average the F1-scores of individual methods within each category and compare the resulting method-category performance across fault types. 
As shown in Fig.~\ref{method_category_classification_f1_lines}, different method categories show similar fault type difficulty trends under the same modality, suggesting that fault-specific characteristics largely determine classification difficulty.
The main exception appears in the TS category on the Metrics modality, where several fault types deviate from this trend.

\begin{figure}[t]
	\centering
	\includegraphics[width=\linewidth]{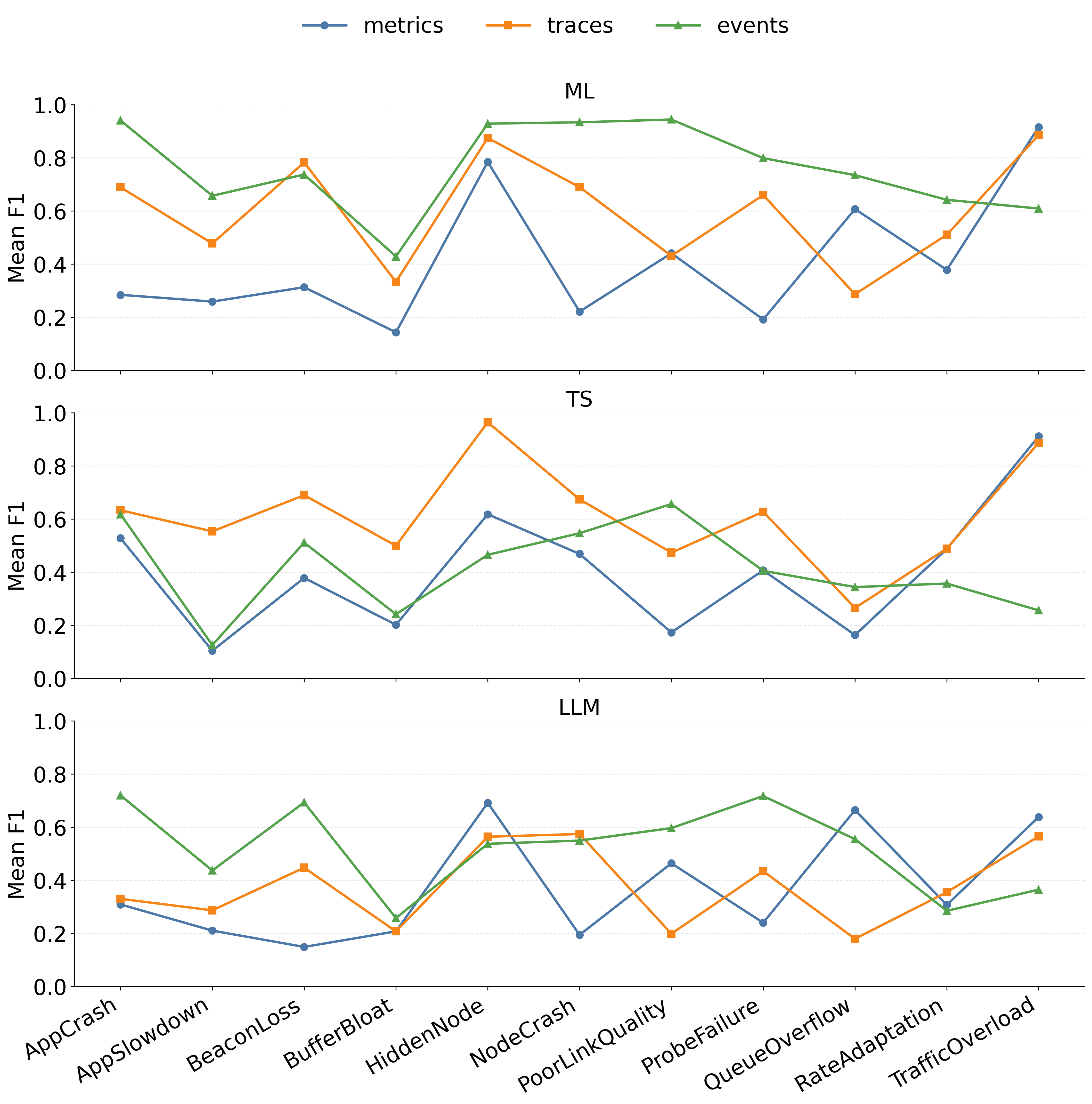}
	\caption{Per-Fault Classification F1 Across Method Modalities and Categories.}
	\label{classification_f1_by_method_modality_lines}
\end{figure}

We further reorganize the fault type level classification results from the perspective of method categories. 
Fig.~\ref{classification_f1_by_method_modality_lines} presents three subfigures for ML, TS, and LLM-based methods, where each curve corresponds to one modality. 
This allows us to compare modality effects across fault types within each method category.
First, although warning events perform well on most fault types, their effectiveness drops on TrafficOverload because this fault is not covered by a corresponding warning event. 
This reveals a key limitation of event-based observations. 
Their high-level abstraction provides strong diagnostic cues when relevant events exist, but also compresses away low-level behavioral details. 
As a result, when a fault is not explicitly represented by predefined events, models can hardly recover the latent fault pattern from the remaining sparse numerical features.
In practice, new event types or rule-based patches can be added to handle emerging faults, but this is not a sustainable solution. 
To avoid such case-by-case extensions, learning fault patterns from raw observations becomes a more attractive direction.

Second, we observe clear modality-dependent performance contrasts between Metrics and Traces for three fault types: PoorLinkQuality, ProbeFailure, and QueueOverflow.
One possible explanation is that traces perform better for faults manifested as disconnection, while Metrics are more effective for faults manifested as latency or stuttering.
For example, ProbeFailure is better captured by Traces, while PoorLinkQuality and QueueOverflow are more effectively reflected by Metrics.
However, this pattern is not absolute, as the curves on the left do not fully follow this tendency. 
This reflects the complexity of Wi-Fi fault behaviors, where explanations of specific modality advantages may not generalize across scenarios. 
Therefore, the key issue is not to assign fixed roles to individual modalities, but to better exploit their complementary diagnostic evidence.

Task-level and fault-level analyses show that the challenge of Wi-Fi fault diagnosis is not merely limited classification accuracy, but the substantial differences in how different faults are observable across modalities.
As a result, aggregate F1-scores can obscure the fine-grained relationships between modalities and fault behaviors.

\subsection{Multi-modal Fusion Analysis}

\begin{table*}[!t]
	\centering
	\caption{Performance improvement under multi-modal fusion.}
	\label{multimodal_fusion}
	\begin{tabular}{ccccc}
		\toprule
		fusion & iperf+tcpdump  & iperf+warning  & tcpdump+warning  & iperf+tcpdump+warning \\
		\midrule
		LR  & +0.05 / +0.09 / +0.11  & +0.06 / +0.06 / +0.09  & +0.07 / +0.09 / +0.09  & +0.07 / +0.09 / +0.09 \\
		MLP  & -0.06 / +0.08 / +0.06  & +0.03 / +0.07 / +0.07  & +0.06 / +0.08 / +0.09  & +0.01 / +0.09 / +0.08 \\
		Attn-MLP  & +0.02 / +0.07 / +0.05  & +0.02 / +0.05 / +0.08  & -0.03 / +0.08 / +0.10  & +0.00 / +0.10 / +0.10 \\
		RF  & +0.02 / +0.05 / +0.04  & -0.02 / +0.03 / -0.05  & +0.09 / +0.06 / +0.00  & +0.10 / +0.07 / -0.00 \\
		SVM  & +0.01 / +0.02 / -0.02  & +0.06 / +0.05 / +0.04  & +0.07 / +0.07 / +0.04  & +0.07 / +0.06 / +0.04 \\
		KNN  & -0.03 / -0.02 / -0.02  & +0.03 / +0.00 / -0.01  & +0.01 / -0.01 / -0.01  & -0.01 / -0.07 / -0.03 \\
		\midrule
		Gemini  & +0.03 / +0.18 / +0.12  & +0.00 / +0.07 / +0.03  & +0.02 / +0.05 / +0.05  & +0.03 / +0.11 / +0.06 \\
		GPT  & +0.02 / +0.13 / +0.06  & +0.00 / -0.02 / +0.01  & -0.00 / +0.01 / +0.06  & +0.04 / +0.02 / +0.02 \\
		LLaMA  & +0.02 / +0.10 / +0.04  & -0.06 / -0.01 / -0.17  & +0.03 / +0.03 / -0.07  & -0.01 / +0.03 / -0.19 \\
		Qwen  & +0.01 / +0.05 / +0.03  & -0.00 / +0.01 / -0.05  & +0.00 / +0.06 / -0.01  & +0.01 / -0.09 / -0.07 \\
		\bottomrule
	\end{tabular}
	\vspace{0.5em}
\end{table*}

\begin{figure}[t]
	\centering
	\includegraphics[width=\linewidth]{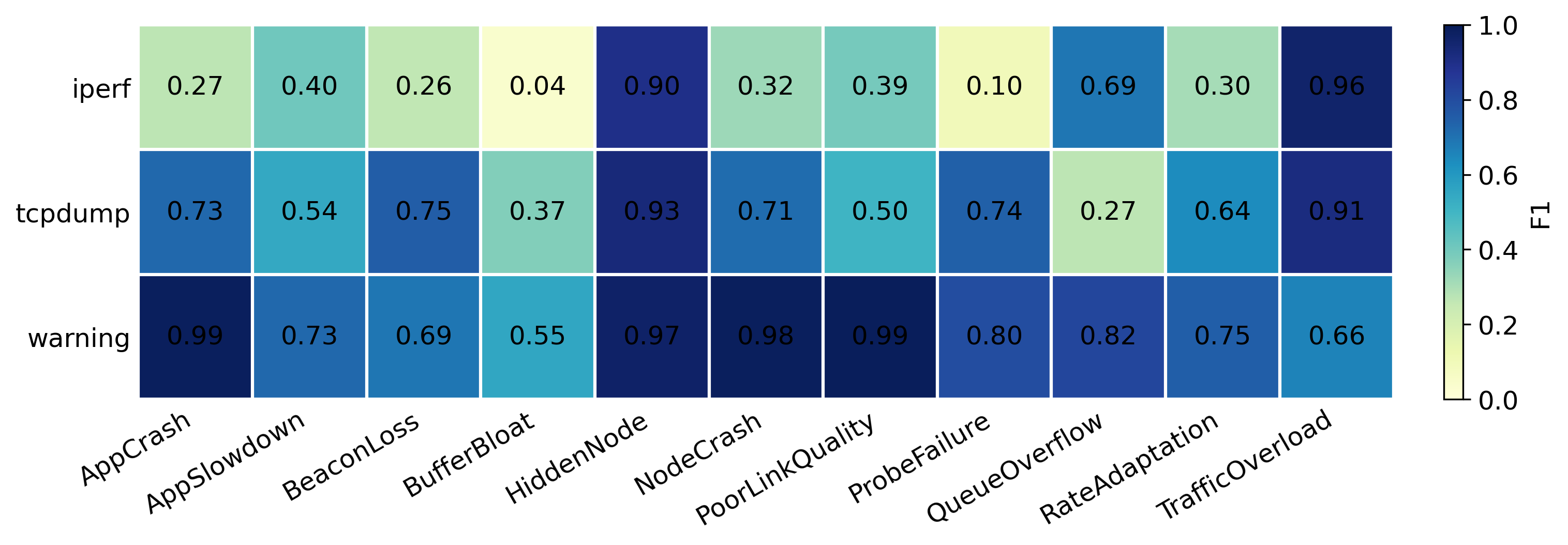}
	\caption{Fault type level classification F1-score under multi-modal fusion.}
	\label{dataset_classification_fault_f1_heatmap}
\end{figure}

We further evaluate multi-modal diagnosis by feeding fused representations into diagnosis algorithms.
Two fusion strategies are reported: simple feature concatenation and attention-based fusion.
The former is applied to all diagnosis algorithms, whereas the latter is evaluated only with MLP due to its learnable neural fusion module.
Table~\ref{multimodal_fusion} shows the improvement of each modality combination over the best single-modality baseline.
The result shows that most modality combinations lead to performance improvements over the best single-modality baseline. 
To further understand where these gains come from, we again decompose the results at the fault type level as shown in Fig~\ref{dataset_classification_fault_f1_heatmap}.
The fault type level results reveal that the gains from fusion are highly concentrated. 
When warning events are included, most of the improvement comes from TrafficOverload. 
Across the remaining fault types, multi-modal fusion brings only marginal gains in most cases and may even reduce performance.

This indicates that the aggregate F1 improvement can be misleading. 
Although adding an additional modality improves the overall score, the actual benefit of multi-modal fusion is concentrated on a limited set of fault types. 
In most cases, fusion appears to smooth the results across modalities and tends to follow the strongest modality, preventing severe degradation but offering limited complementary gains.
The attention-based fusion mechanism mainly provides adaptive weighting over modalities. 
However, adaptive weighting alone may be insufficient to capture complex cross-modal interactions.

LLM-based methods exhibit more complex behavior, as different LLMs operate as black-box systems with model-specific reasoning and interpretation capabilities. 
Among them, Gemini shows the strongest improvement under multi-modal inputs. 
Notably, its relative gain is larger than that of traditional methods, although its absolute performance remains lower.

Multi-modal fusion improves overall performance, but its benefits are uneven and fault-dependent. Current fusion strategies mainly stabilize performance across modalities rather than consistently exploiting complementary evidence across faults.

\subsection{Reasoning-oriented Evaluation Results}

To validate the effectiveness of the proposed operational feature representation, we further evaluate its performance on downstream fault classification tasks. Table~\ref{Operational Feature Classification} summarizes the fault classification results using operational feature representations as model inputs.

\begin{table}[!t]
	\centering
	\caption{Fault Classification Performance Using Operational Feature Representations.}
	\label{Operational Feature Classification}
	\begin{tabular}{cccccc}
		\toprule
		task/model & mlp & RF & LR & SVM & KNN \\
		\midrule
		detection & 0.997 & 0.999 & 0.975 & 0.998 & 0.949 \\
		classification & 0.905 & 0.914 & 0.798 & 0.921 & 0.845 \\
		localization & 0.990 & 0.997 & 0.988 & 0.993 & 0.973 \\
		\bottomrule
	\end{tabular}
	\vspace{0.5em}
\end{table}

Compared with all single-modality diagnosis results, operational feature representations achieve substantially stronger fault classification performance across different downstream classifiers. This indicates that the constructed operational features remain highly correlated with underlying fault conditions and can effectively preserve fault-related operational characteristics for downstream diagnosis tasks.

\begin{figure}[t]
	\centering
	\includegraphics[width=\linewidth]{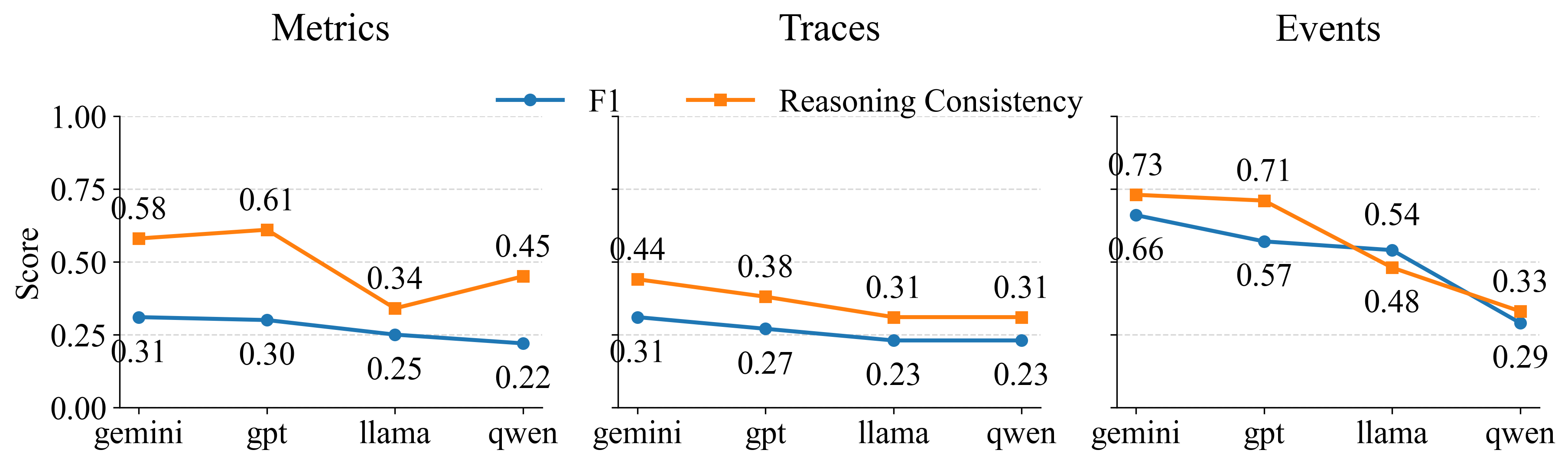}
	\caption{Performance and Reasoning Consistency of LLMs Across Modalities}
	\label{llm_f1_causal_curve}
\end{figure}

We further evaluate diagnostic reasoning consistency between LLM-generated operational feature representations and ground-truth operational features.Since the reasoning-oriented evaluation is conducted only on the original raw samples before distillation, the dataset size remains sufficiently small to allow exhaustive threshold search for optimal feature activation calibration. Specifically, the threshold values for operational feature activation are obtained through exhaustive enumeration to maximize the reasoning consistency F1-score on the evaluation set. The resulting reasoning consistency performance shown in Fig~\ref{llm_f1_causal_curve}.
The orange line represents reasoning consistency, while the blue line represents the F1 score.
The two curves exhibit broadly similar trends, indicating that the proposed reasoning consistency metric is generally aligned with downstream diagnostic performance.

Although the reasoning consistency results remain relatively strong across different observation modalities, downstream fault classification performance remains substantially lower. This phenomenon suggests that current LLM-based approaches can partially understand coarse-grained operational network behaviors and abnormal events, while still struggling to reliably infer specific fault categories from complex quantitative relationships among heterogeneous network observations.

The experimental results suggest two important limitations of current LLM-based diagnosis approaches under realistic wireless environments. First, current LLMs still lack reliable domain-specific reasoning relationships between operational network conditions and underlying fault categories. We further observe that carefully designed reasoning pipelines can significantly improve diagnosis performance for faults with relatively explicit causal structures, such as application crashes, where deterministic operational behaviors can be directly associated with fault conditions. Under such manually designed reasoning procedures, several strongly causal fault categories can achieve nearly perfect diagnosis accuracy. These observations indicate that part of the diagnosis limitation originates from insufficient network-domain reasoning knowledge rather than pure reasoning capability itself.

\begin{table}[!t]
	\centering
	\caption{Reasoning consistency improvement under multi-modal.}
	\label{Multi_model_reasoning}
	\begin{tabular}{ccccc}
		\toprule
		model
		& \makecell{iperf + \\ tcpdump} 
		& \makecell{iperf + \\ warning} 
		& \makecell{tcpdump + \\ warning}
		& all \\
		\midrule
		
		Gemini  & -0.07  & -0.05  & -0.05  & -0.05 \\
		GPT  & -0.12  & -0.04  & -0.20  & -0.10 \\
		LLaMA  & -0.07  & -0.07  & -0.07  & -0.09 \\
		Qwen  & -0.18  & -0.15  & -0.08  & -0.21 \\
		\bottomrule
	\end{tabular}
	\vspace{0.5em}
\end{table}

However, relying solely on prior knowledge is also unreliable.
For example, for faults such as Bufferbloat, the concrete observable symptoms are difficult to specify precisely in advance, making knowledge-based judgment criteria prone to failure when the actual manifestation differs from the predefined description.
This further leads to the second and more fundamentally challenge, current LLMs remain insufficiently sensitive to fine-grained quantitative operational variations, even when numerical observations are transformed into discretized operational states. Certain faults, such as buffer bloat, often manifest through distributed quantitative influences across multiple correlated network indicators rather than explicit abnormal operational events. In such cases, fault-related behaviors are implicitly reflected through complex interactions among delay, jitter, retransmissions, and throughput variations, making it difficult to construct explicit rule-based reasoning mechanisms with strong generalization capability.
In contrast, conventional neural network models can often learn such hidden quantitative correlations directly from large-scale statistical observations. 

We investigate the reasoning consistency of LLMs under multi-modal diagnostic inputs, with the results reported in Table~\ref{Multi_model_reasoning}.
An interesting discrepancy can be observed from the multi-modal results.
Across all multi-modal settings, reasoning consistency decreases compared with the corresponding single-modality settings.
In contrast, the downstream LLM-based diagnostic performance generally improves, as shown in Table~\ref{multimodal_fusion}.
This result suggests that multi-modal observations may introduce additional interference into the LLM's interpretation of network states.
Due to the lack of sufficient cross-modal correlation knowledge, the LLM may struggle to effectively identify and integrate useful evidence from heterogeneous observations.
However, the additional modalities can still be reflected in the extracted feature representations, and such differences may be captured by the downstream classifier, leading to improved prediction accuracy.
This indicates that multi-modal data introduce representation-level variations or complementary signals within the LLM, but do not lead to stable and interpretable cross-modal diagnostic understanding.

These results show that LLMs can provide useful intermediate representations for downstream diagnosis, but their generated analyses do not yet demonstrate stable Wi-Fi fault understanding.
This gap between prediction utility and reasoning consistency highlights the need to evaluate LLM-based diagnosis beyond final accuracy.

\subsection{Discussion}

Overall, our results show that different modalities play different roles in Wi-Fi fault diagnosis.
Warning events provide strong diagnostic cues, but they should not be viewed in the same way as general-purpose observations such as metrics and packet traces.
Since warning events are high-level abstractions designed to reflect fault-related network states, directly using them as diagnosis inputs may overestimate the achievable performance of event-based modalities.
In this sense, warning events are more appropriate as structured reflections of the underlying network conditions or as reference evidence for reasoning-oriented evaluation.

Although multi-modal data can improve overall diagnostic performance, aggregate F1-scores may obscure the actual source of the improvement.
For fault classification, decomposing the results by fault type provides a more informative view of how each modality affects specific faults.
This analysis shows whether multi-modal fusion provides general complementary benefits or only improves a limited subset of faults, thereby helping reveal the weaknesses of different modalities in concrete fault scenarios.
For example, although simple modality concatenation improves the aggregate F1-score, the fault type level analysis shows that this improvement can be highly uneven.
The gain may mainly come from a large improvement on one difficult fault type, while the performance on most other faults is marginally improved or even degraded.
This makes it difficult to conclude that simple concatenation provides an effective multi-modal fusion mechanism.
A similar observation can be made for attention-based fusion.
While attention-based fusion improves the overall F1-score, it may degrade the performance of fault types that already achieve strong single-modality results.
This poses a challenge for further improving diagnosis performance.

LLM-based methods do not show particularly strong diagnostic performance in our benchmark.
Their main advantage lies in the inspectability of their generated outputs, which provides an opportunity to analyze where the diagnostic process fails.
By examining the LLM-generated analyses and evidence representations, we find that improving LLM-based fault diagnosis requires more comprehensive network-domain knowledge and additional mechanisms for handling numerical observations.
Specifically, current LLMs still struggle to associate heterogeneous operational symptoms with underlying fault causes, while also remaining insufficiently sensitive to fine-grained quantitative variations.

\section{Conclusion}

In this paper, we studied realistic Wi-Fi fault diagnosis from a multi-modal benchmarking perspective.
Motivated by the lack of public operational data and unified evaluation settings for Wi-Fi fault diagnosis, we constructed a real-world Wi-Fi fault diagnosis testbed with automated fault injection and synchronized cross-layer data collection.
The resulting dataset provides a common basis for examining how Wi-Fi faults are reflected across heterogeneous operational observations, including performance metrics, packet traces, warning events, and monitor logs.

Based on this benchmark, our study shows that multi-modal Wi-Fi fault diagnosis remains a non-trivial problem.
Multi-modal observations contain useful diagnostic information, but existing methods still struggle to fully exploit their potential.
Although adding more modalities can improve overall performance, the gains are often uneven across tasks and fault types, indicating that current fusion strategies have not yet effectively captured the relationships among heterogeneous Wi-Fi fault observations.
Our analysis further shows that current LLM-based approaches provide new opportunities for inspectable diagnostic analysis, but still lack sufficient network-domain reasoning ability and numerical sensitivity for reliable Wi-Fi fault diagnosis.
These findings suggest that future work should not only collect richer operational modalities, but also develop methods that can identify useful evidence from heterogeneous observations and determine how different modalities contribute to specific Wi-Fi faults.

\bibliography{refs}

\begin{IEEEbiography}[{\includegraphics[width=1in,height=1.25in,clip,keepaspectratio]{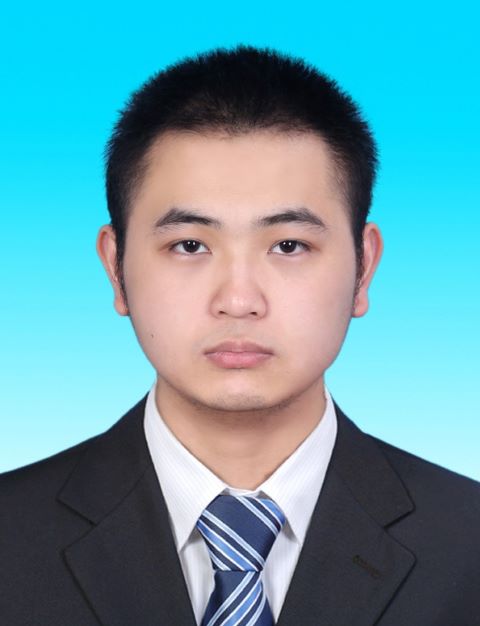}}]{Junjian Zhang}
	received the B.S. degree from Hunan University, Changsha, China, in 2015, 
	and the M.S. degree from Huazhong University of Science and Technology, Wuhan, China, in 2018. 
	He is currently pursuing the Ph.D. degree at Central South University, Changsha, China. 
	His research interests include digital twin networks, network communication, and network fault diagnosis.
\end{IEEEbiography}

\begin{IEEEbiography}[{\includegraphics[width=1in,height=1.25in,clip,keepaspectratio]{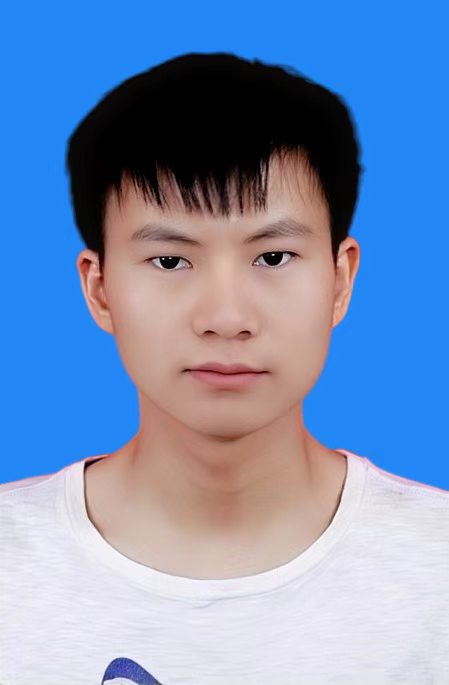}}]{Haobo Deng}
	is currently pursuing the B.S. degree in Software Engineering at Central South University, Changsha, China.  
	His research interests include network fault diagnosis and large language models.
\end{IEEEbiography}

\begin{IEEEbiography}[{\includegraphics[width=1in,height=1.25in,clip,keepaspectratio]{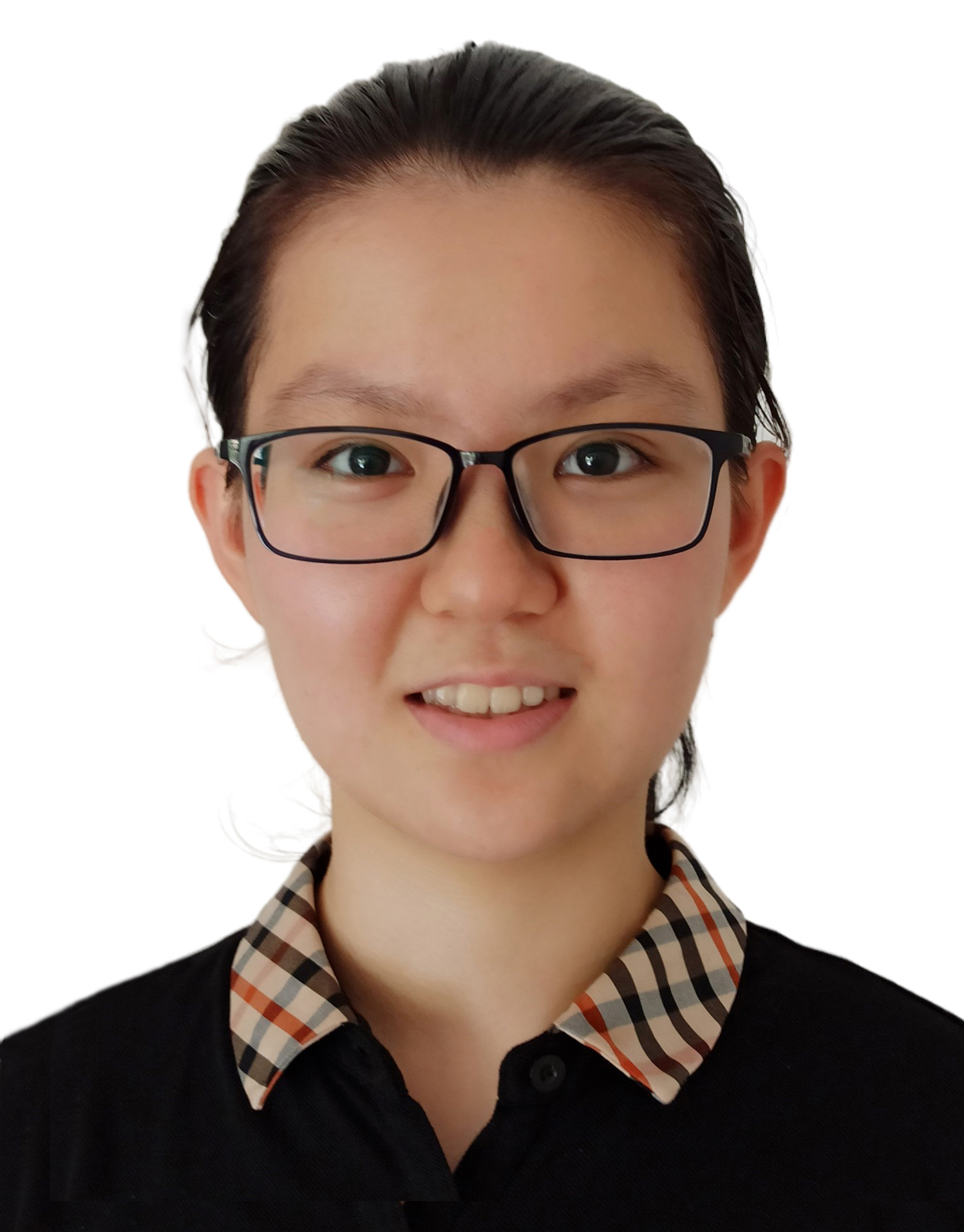}}]{Xinxin Li}
	is currently pursuing the B.S. degree in Software Engineering at Central South University, Changsha, China.
	Her research interests include software engineering and artificial intelligence.
\end{IEEEbiography}

\begin{IEEEbiography}
	[{\includegraphics[width=1in,height=1.25in,clip,keepaspectratio]
		{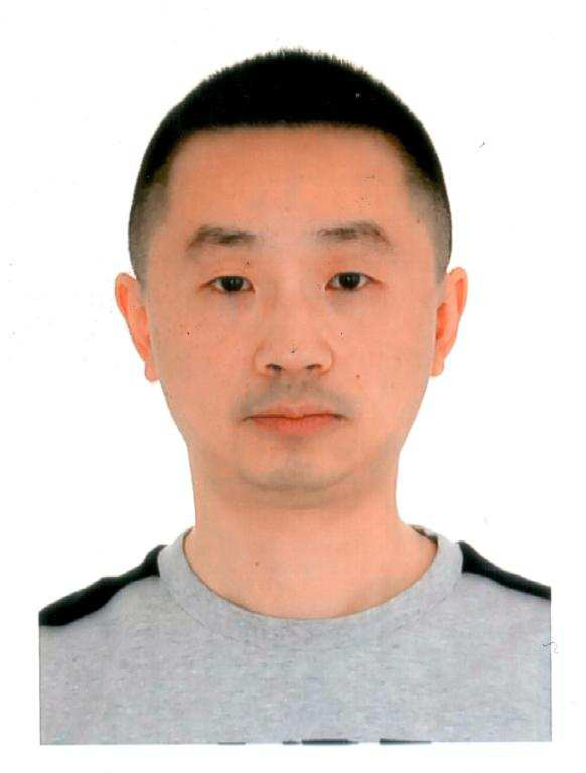}}]{Ming Zhao} (Member, IEEE) received the M.Sc. and Ph.D. degrees in computer science from Central South University, Changsha, China, in 2003 and 2007, respectively. He is currently a Professor with the School of Computer Science and Engineering, Central South University. His main research focuses on wireless networks. He is also a Member of the China Computer Federation.
\end{IEEEbiography}

\begin{IEEEbiography}[{\includegraphics[width=1in,height=1.25in,clip,keepaspectratio]{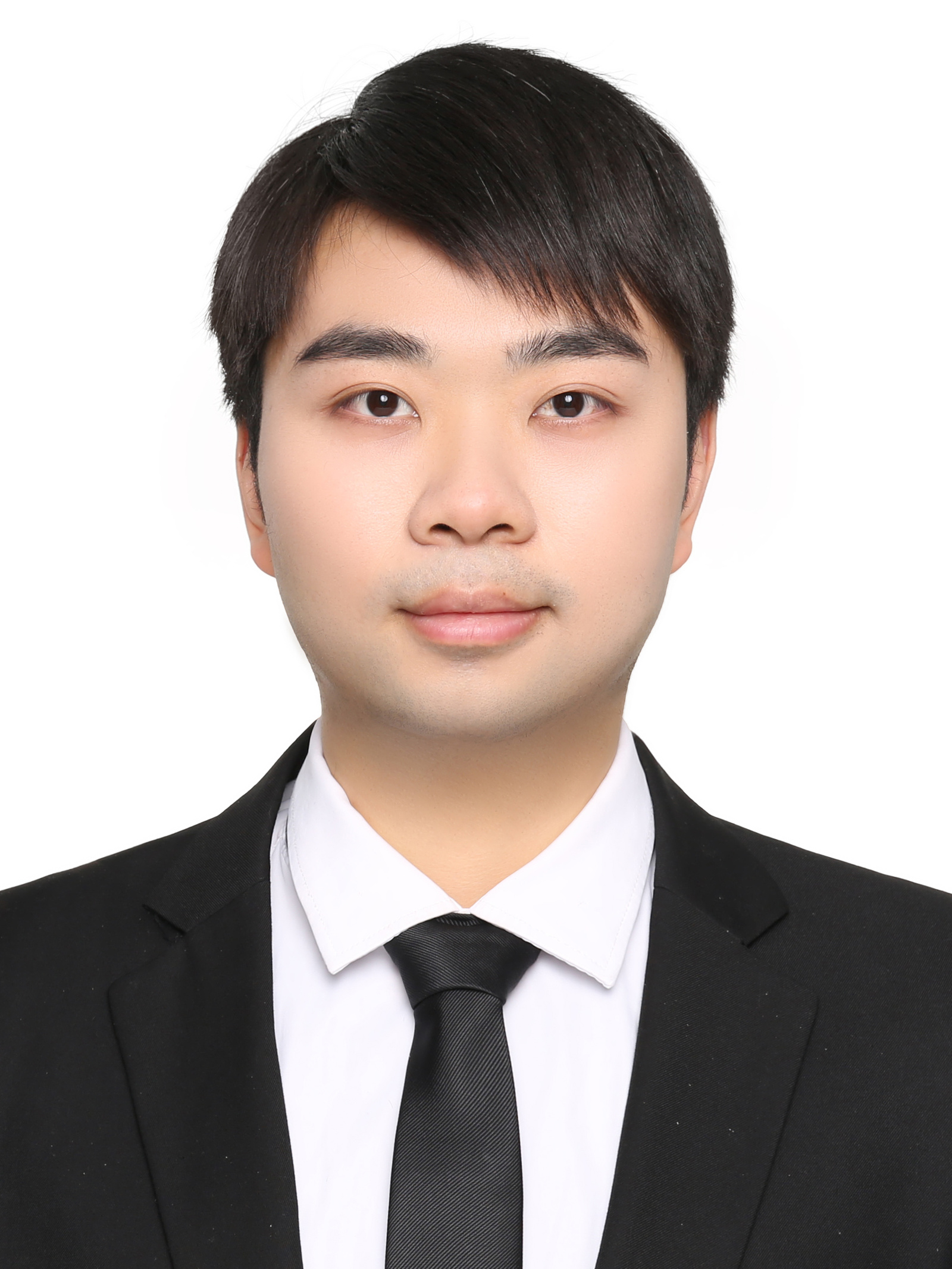}}]{Fengxiao Tang} (Senior Member, IEEE) is a full professor with School of Computer Science and Engineering, Central South University and a distinguished professor in the Graduate School of Information Sciences(GSIS), Tohoku University. He has been an Assistant Professor from 2019 to 2020 and an Associate Professor from 2020 to 2021 at the Graduate School of Information Sciences (GSIS) of Tohoku University. His research interests are unmanned aerial vehicles system, IoT security, game theory optimization, network traffic control and machine learning algorithm. He was a recipient of the prestigious Dean's and President's Awards from Tohoku University in 2019, and several best paper awards at conferences including IC-NIDC 2018/2023, GLOBECOM 2017/2018. He was also a recipient of the prestigious Funai Research Award in 2020, IEEE ComSoc Asia-Pacific (AP) Outstanding Paper Award in 2020 and IEEE ComSoc AP Outstanding Young Researcher Award in 2021.
	
\end{IEEEbiography}

\begin{IEEEbiography}[{\includegraphics[width=1in,height=1.25in,clip,keepaspectratio]{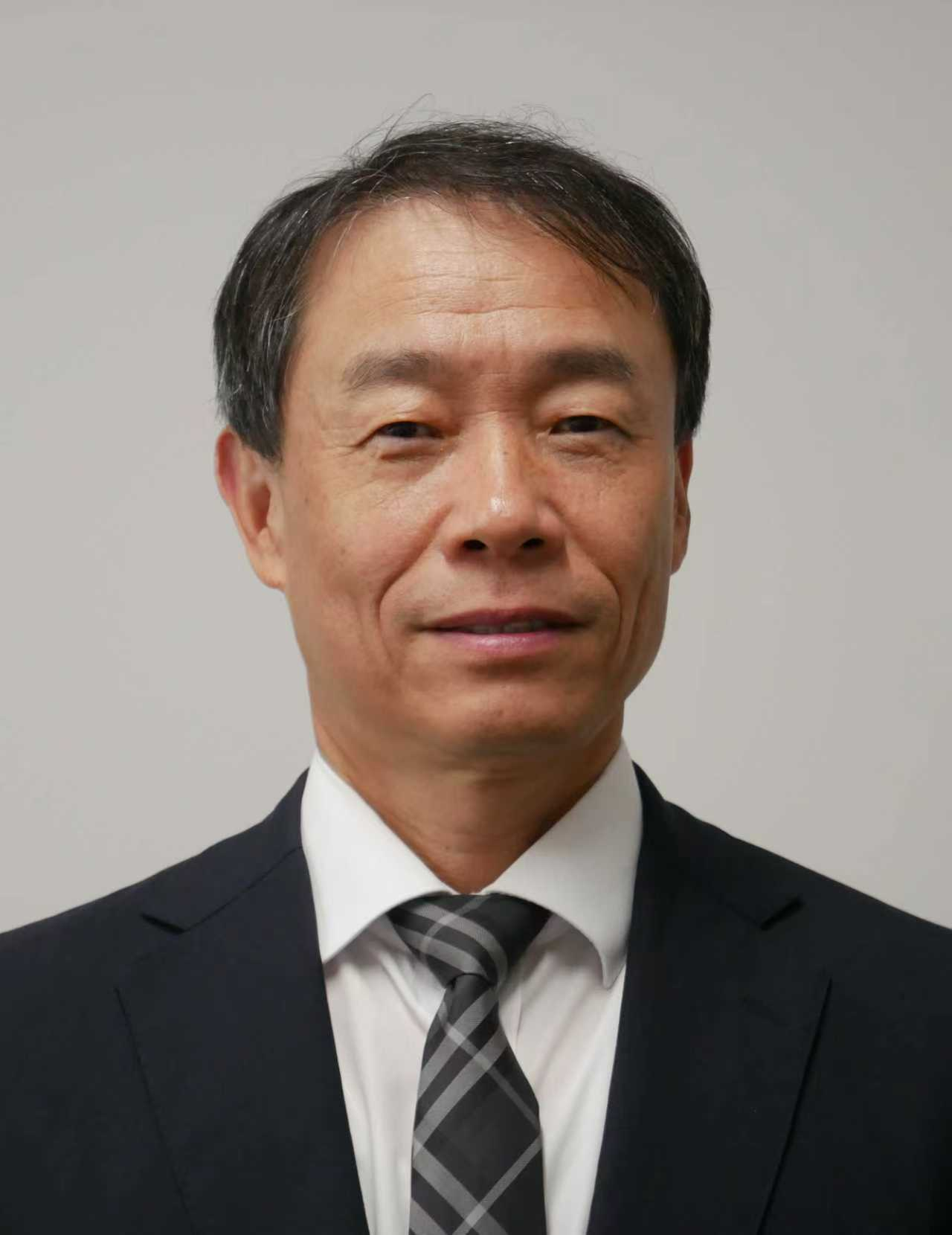}}]{Nei Kato} (Fellow, IEEE)
is a Distinguished Professor with Graduate School of Information Sciences, Tohoku University. He served as the Dean of Graduate School of Information Sciences from 2021 to 2025. His research areas include computer networking, wireless mobile communications, satellite communications, ad hoc \& sensor \& mesh networks, UAV networks, AI, IoT, and Big Data. He is the Editor-in-Chief of IEEE Internet of Things Journal, the Vice President for publication, IEEE Communications Society. He is a Clarivate Analytics Highly Cited Researcher, a Fellow of the Engineering Academy of Japan, a Fellow of IEEE, and a Fellow of IEICE.

\end{IEEEbiography}

\vfill

\end{document}